\documentclass[a4paper,11pt]{article}
\pdfoutput=1

\usepackage{jheppub} 
\usepackage[numbers]{natbib}
\usepackage{braket}
\usepackage{parskip}
\usepackage{mathtools}
\usepackage{bm}
\usepackage{graphbox}
\usepackage{amssymb}
\usepackage{enumitem}
\usepackage{pict2e}
\usepackage{wrapfig}
\usepackage{caption}

\DeclareCaptionLabelSeparator{none}{ }
\hypersetup{colorlinks=true, linkcolor=magenta, citecolor=magenta, urlcolor=magenta}

\makeatletter
\renewcommand{\@makefntext}[1]{%
  \noindent
  \makebox[1.5em][l]{\@thefnmark.}
  \parbox[t]{\dimexpr\linewidth-1.8em}{#1}
}
\makeatother

\newcommand\one{\unitlength=1pt
 \hspace{2.5pt}\begin{picture}(3,1)
 \linethickness{0.14mm}
 \put(-0.7,0.2){\line(0,1){6.5}}
 \put(0.7,0.2){\line(0,1){7.2}}
 \put(-2.1,0.2){\line(1,0){4.2}}
 \put(0.9,7.4){\line(-1,-0.4){3.2}}
 \end{picture}}

\title{On Unitary 2-Group Symmetries}

\author{Thomas Bartsch}
\affiliation{Department of Mathematical Sciences, Durham University, \\
Upper Mountjoy, Stockton Road, Durham, DH1 3LE, \\
United Kingdom}
\emailAdd{thomas.d.bartsch@durham.ac.uk}

\abstract{Global internal symmetries act unitarily on local observables or states of a quantum system. In this note, we aim to generalise this statement to extended observables by considering unitary actions of finite global 2-group symmetries $\mathcal{G}$ on line operators. We propose that the latter transform in unitary 2-representations of $\mathcal{G}$, which we classify up to unitary equivalence. Our results recover the known classification of ordinary 2-representations of finite 2-groups, but provide additional data interpreted as a type of reflection anomaly for $\mathcal{G}$.}

\begin{document} 
\maketitle
\flushbottom

\section{Introduction}

According to Wigner's theorem, invertible global symmetries act unitarily (or anti-unitarily) on local observables or states of a quantum system \cite{Wigner1931}. It is natural to ask how this statement generalises to extended observables. In this note, we try to answer this question by studying unitary actions of a finite global symmetry 2-group $\mathcal{G}$ on line operators in quantum field theory. We propose that while local operators transform in unitary representations of the 0-form part $G \subset \mathcal{G}$, line operators transform in (an appropriate notion of) unitary 2-representations of $\mathcal{G}$, which we describe and classify in this paper.

\subsection{Motivation}

In Euclidean (Wick-rotated) quantum field theory, unitarity manifests itself in the principle of \textit{reflection positivity} \cite{Jaffe:2018ftu,Freed:2016rqq}. Concretely, upon fixing an affine hyperplane $\Pi$ in $D$-dimensional spacetime, the \textit{reflection} part implies that reflecting the operator content of a correlation function about $\Pi$ is equivalent to complex conjugating the correlation function,
\begin{equation}
\vspace{-5pt}
\begin{gathered}
\includegraphics[height=1.75cm]{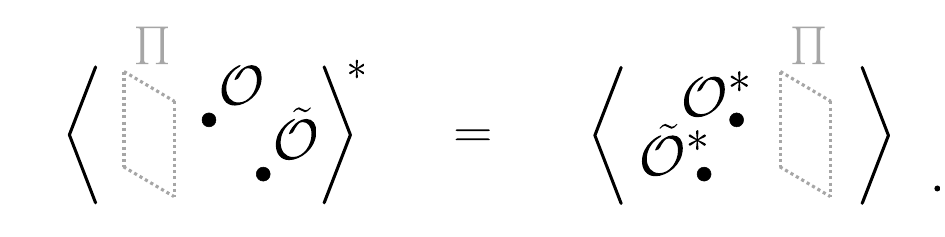}
\end{gathered}
\end{equation}
Here, $\mathcal{O}^{\ast}$ denotes the operator that is obtained by reflecting a given operator $\mathcal{O}$ about the fixed hyperplane $\Pi$. As a special case, we can consider the half-space correlation function\footnote{Here, we use an operator-state map which maps a local operator to a state in the Hilbert space of the theory using standard path integral methods. Since we don't assume the theory to be conformal, this map is not surjective in general.}
\begin{equation}
\begin{gathered}
\includegraphics[height=1.75cm]{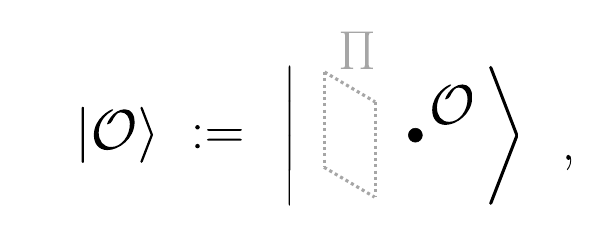}
\end{gathered}
\end{equation}
which is a vector in the Hilbert space $\mathcal{H}$ of the theory \cite{Freed:2016rqq}. Reflecting about $\Pi$ then gives a vector in the complex conjugate Hilbert space $\mathcal{H}^{\ast}$,
\begin{equation}
\begin{gathered}
\includegraphics[height=1.75cm]{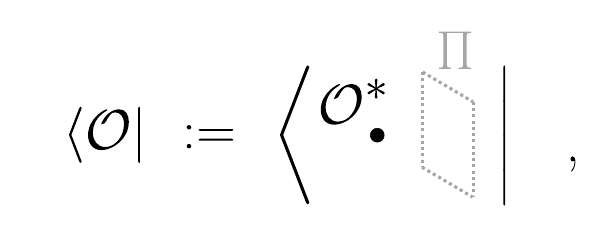}
\end{gathered}
\end{equation}
which is canonically identified with a linear functional $\bra{\mathcal{O}} \in \mathcal{H}^{\vee}$ in the dual space of $\mathcal{H}$. This then allows us to define overlaps
\begin{equation}
\begin{gathered}
\includegraphics[height=1.75cm]{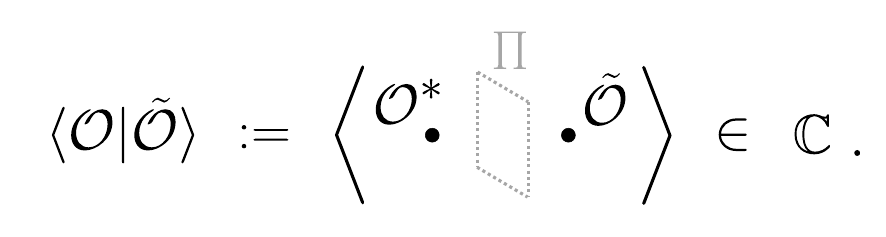}
\end{gathered}
\end{equation}
\textit{Positivity} is the statement that these overlaps are positive definite, i.e. $\braket{\mathcal{O} | \mathcal{O}} \geq 0$ for all $\ket{\mathcal{O}} \in \mathcal{H}$ with equality if and only if $\ket{\mathcal{O}} = 0$.

Now suppose that the quantum field theory admits a finite global symmetry group $G$, which is implemented by codimension-one topological defects labelled by group elements $g \in G$ that fuse according to the group law of $G$:
\begin{equation}
\vspace{-5pt}
\begin{gathered}
\includegraphics[height=2.5cm]{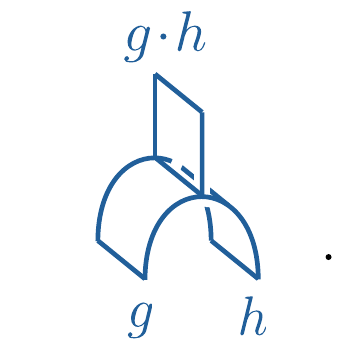}
\end{gathered}
\end{equation}
The symmetry group $G$ can then act on local operators via linking \cite{Gaiotto:2014kfa}, i.e.
\begin{equation}
\begin{gathered}
\includegraphics[height=1.4cm]{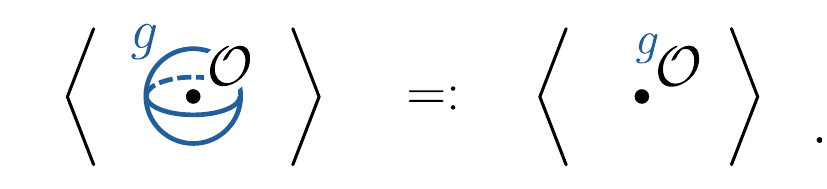}
\end{gathered}
\end{equation}
Equivalently, we can define an action of group elements $g \in G$ on states $\ket{\mathcal{O}} \in \mathcal{H}$ via
\begin{equation}
\begin{gathered}
\includegraphics[height=1.75cm]{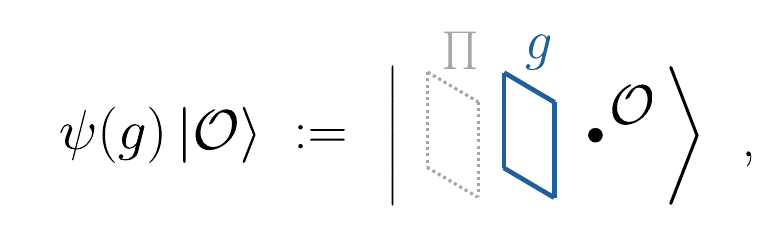}
\end{gathered}
\end{equation}
furnishing a representation $\psi$ of $G$ on the Hilbert space $\mathcal{H}$. Using the fact that reflection about $\Pi$ acts by $\ast: g \mapsto g^{-1}$ on the topological symmetry defects $g \in G$, we then have that
\begin{equation}
\begin{gathered}
\includegraphics[height=3.5cm]{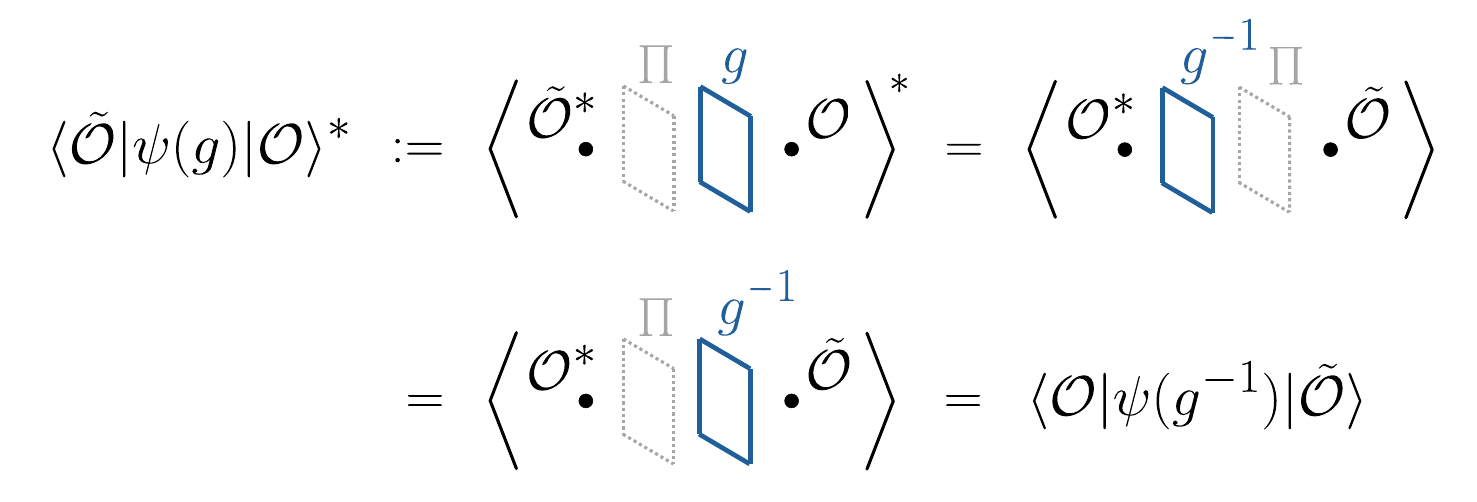}
\end{gathered}
\end{equation}
for all $\mathcal{O}$ and $\tilde{\mathcal{O}}$, which implies that $\psi(g)^{\dagger} = \psi(g^{-1})$ for all $g \in G$. Hence, we see that the representation $\psi$ of $G$ on $\mathcal{H}$ is unitary.

In spacetime dimension $D > 2$, global symmetries can also act on extended operators such as line operators \cite{Delmastro:2022pfo,Bhardwaj:2023wzd,Bartsch:2023pzl,Bartsch2023a,Bhardwaj:2023ayw}. In the following, we assume that all line operators $L$ are \textit{simple} in the sense that they only host topological local operators proportional to the identity $\text{id}_L$ on $L$. Given such a line operator $L$, the symmetry group $G$ can act on it via wrapping, i.e.
\begin{equation}
\vspace{-5pt}
\begin{gathered}
\includegraphics[height=1.45cm]{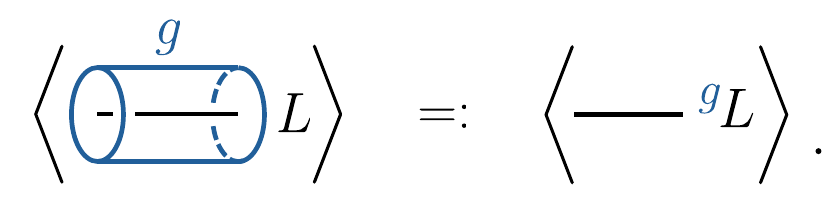}
\end{gathered}
\end{equation}
Equivalently, ${}^gL$ is the unique line operator such that there exists a one-dimensional space of local intersection operators 
\begin{equation}
\label{eq-2-rep-intersection-operator}
\vspace{-5pt}
\begin{gathered}
\includegraphics[height=1.5cm]{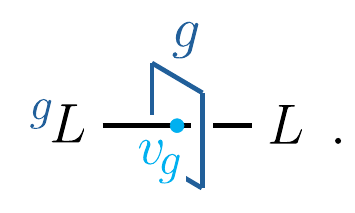}
\end{gathered}
\end{equation}
Without loss of generality, we now assume that the line $L$ is fixed by the whole of $G$, i.e. ${}^gL = L$ for all $g \in G$ (the more general case of a proper stabiliser subgroup $H \subset G$ can be obtained by induction). We then fix for each $g \in G$ a local intersection operator $v_g$ as in (\ref{eq-2-rep-intersection-operator}) such that $v_e = \text{id}_L$ (where $e \in G$ is the identity element in $G$). The action of group elements $g,h \in H$ on $L$ may then carry an 't Hooft anomaly in the sense that
\begin{equation}
\vspace{-5pt}
\begin{gathered}
\includegraphics[height=1.7cm]{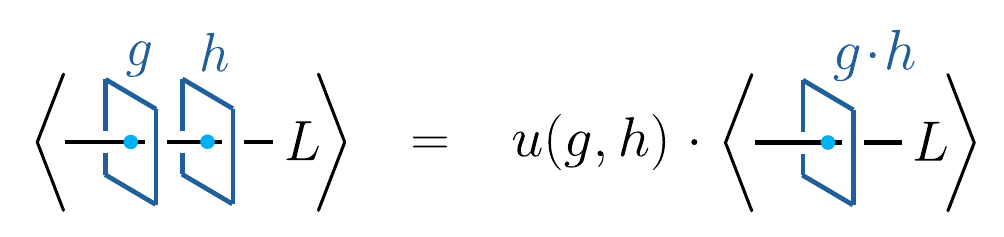}
\end{gathered}
\end{equation}
for some multiplicative phase $u(g,h) \in U(1)$, which corresponds to the composition law
\begin{equation}
\label{eq-intersection-composition-law}
v_g \circ v_h \; = \; u(g,h) \cdot v_{gh}
\end{equation}
for the local intersection operators $v_g$. In order for this to be compatible with associativity of the group multiplication in $G$, the collection of phases $u(g,h)$ needs to define a (normalised) 2-cocycle $u \in Z^2(G,U(1))$. Similarly, reflecting the intersection operators $v_g$ about $\Pi$ may produce anomalous phases
\begin{equation}
\vspace{-5pt}
\begin{gathered}
\includegraphics[height=1.68cm]{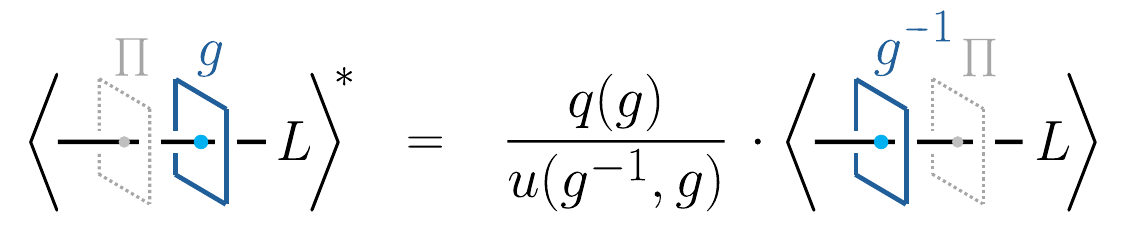}
\end{gathered}
\end{equation}
for some $q(g) \in U(1)$, which corresponds to the reflection law
\begin{equation}
(v_g)^{\ast} \; = \; \frac{q(g)}{u(g^{-1},g)} \hspace{1pt}\cdot\hspace{1pt} v_{(g^{-1})} \; .
\end{equation}
In order for this to be compatible with the involutariness of the reflection $\ast$ as well as the composition law (\ref{eq-intersection-composition-law}), $q$ needs to define a group homomorphism
\begin{equation}
q \; \in \; \text{Hom}(G,\mathbb{Z}_2) \; ,
\end{equation}
which we interpret as a type of \textit{reflection anomaly} for $G$ on the line operator $L$.

In order to see further implications of this reflection anomaly, we assume that the line $L$ can end on twisted sector local operators
\begin{equation}
\vspace{-5pt}
\begin{gathered}
\includegraphics[height=1.1cm]{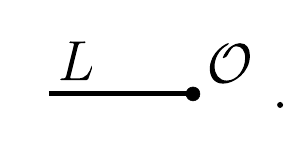}
\end{gathered}
\end{equation}
As before, we can construct half-space correlation functions 
\begin{equation}
\begin{gathered}
\includegraphics[height=1.7cm]{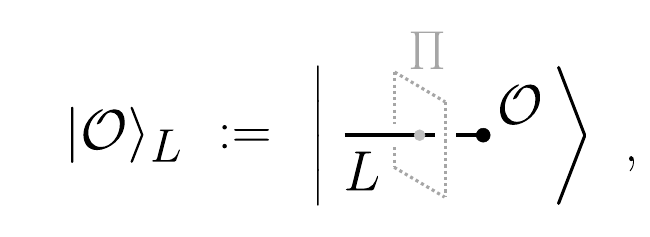}
\end{gathered}
\end{equation}
which correspond to states in the $L$-twisted Hilbert space $\mathcal{H}_L$. The symmetry group $G$ then acts on these states via
\begin{equation}
\begin{gathered}
\includegraphics[height=1.7cm]{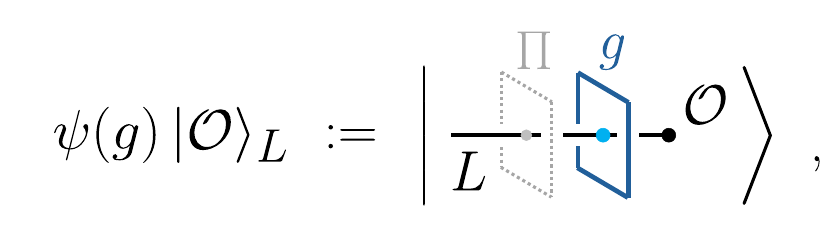}
\end{gathered}
\end{equation}
where the linear maps $\psi(g)$ satisfy the composition law
\begin{equation}
\psi(g) \,\circ \, \psi(h) \; = \; u(g,h) \hspace{1pt}\cdot\hspace{1pt} \psi(gh) \; .
\end{equation}
Moreover, by considering overlaps of twisted sector states, we find that
\begin{equation}
\begin{gathered}
\includegraphics[height=3.5cm]{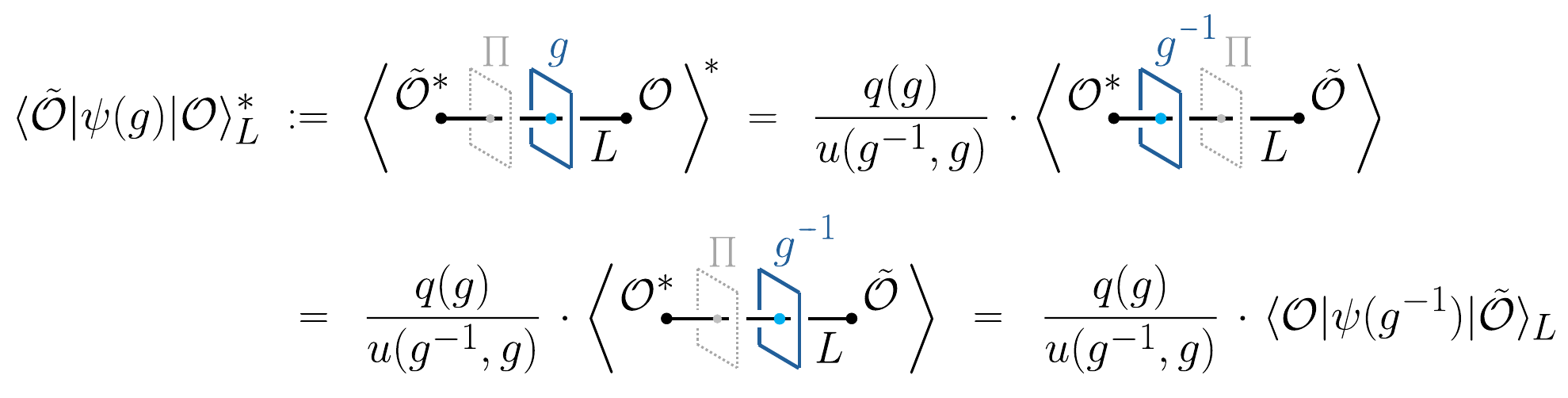}
\end{gathered}
\end{equation}
for all $\mathcal{O}$ and $\tilde{\mathcal{O}}$, which implies that, as operators on $\mathcal{H}_L$, we have
\begin{equation}
\psi(g)^{\dagger} \,\; = \,\; \frac{q(g)}{u(g^{-1},g)} \cdot \psi(g^{-1}) \,\; \equiv \,\; q(g) \cdot \psi(g)^{-1}
\end{equation}
for all $g \in G$. As a result, the norm squared of a state $\psi(g) \ket{\mathcal{O}} \in \mathcal{H}_L$ is given by
\begin{equation}
\big\lVert \psi(g) \ket{\mathcal{O}} \hspace{-1pt}\big\rVert^2_L \,\; = \,\; q(g) \cdot \big\lVert \ket{\mathcal{O}} \big\rVert_L^2 \; ,
\end{equation}
which, in order to be compatible with positivity of $\mathcal{H}_L$, requires $q(g) = 1$ for all $g \in G$. Hence, we see that a necessary condition for the line operator $L$ to be able to end on twisted sector local operators is that the associated reflection anomaly $q$ vanishes. If this is the case, $\psi$ defines a unitary representation of $G$ on $\mathcal{H}_L$ with projective 2-cocycle $u$. 

All in all, we see that local operators (genuine or twisted) transform in unitary representations of the global symmetry group $G$. The aim of this note is to provide an analogous statement for the action of $G$ on line operators, by identifying the tuple $(u,q)$ with a (certain type of) unitary 2-representation of $G$. This will generalise the notion of unitary actions of global symmetry groups from local to extended operators.

\subsection{Summary}

Local operators in a unitary quantum field theory form a Hilbert space, which a finite global symmetry group $G$ acts on via unitary representations. Since according to Maschke's theorem every such representation of $G$ is a direct sum of irreducible ones, we may without loss of generality assume that the above Hilbert space is finite-dimensional\footnote{Unless stated otherwise, all vector spaces in this note will be finite-dimensional complex vector spaces. We denote the category of such vector spaces and linear maps between them by $\text{Vect}$ in what follows.}. A unitary representation of $G$ then corresponds to a $\dagger$-functor \cite{Selinger:2007eep,stehouwer2023}
\begin{equation}
\label{eq-positive-unitary-1-rep}
\psi: \; BG \; \to \; \text{Hilb}
\end{equation}
from the delooping of $G$ into the category $\text{Hilb}$ of finite-dimensional complex Hilbert spaces. In this note, we aim to generalise the above to the action of $G$ on extended line operators by making the following propositions:
\begin{enumerate}
\item \parbox[t]{\dimexpr\textwidth-\leftmargin}{\vspace{-2.5mm}
\begin{wrapfigure}[4]{r}{0.1496\textwidth}
  \vspace{-10pt}
  \begin{center}
    \vspace{-2pt}
    \includegraphics[height=1.05cm]{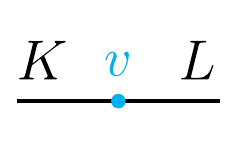}
    \vspace{-7pt}
    \caption{}
    \label{fig-line-junctions}
  \end{center}
\end{wrapfigure}
We propose that line operators in a unitary quantum field theory form a 2-Hilbert space in the sense of \cite{BAEZ1997125} (see also \citep{Freed:1994ad,Bartlett:2008ji,Chen:2024pnk}), which captures the category of line operators and topological junctions between them as illustrated in Figure \ref{fig-line-junctions}.}
\vspace{3pt}
\item We propose that the symmetry group $G$ acts on line operators via unitary 2-represen-tations, which correspond to $\dagger$-2-functors
\begin{equation}
\rho: \; BG \; \to \; \text{2Hilb}
\end{equation}
from the delooping of $G$ into the 2-category $\text{2Hilb}$ of 2-Hilbert spaces. This requires introducing a notion of higher $\dagger$-categories and higher $\dagger$-functors between them. As described in \cite{Ferrer:2024vpn}, there is a variety of flavours of $\dagger$-2-categories corresponding to different choices of subgroup $\mathfrak{G} \subset \text{Aut}(\text{Cat}_{(\infty,2)}) = (\mathbb{Z}_2)^2$ of the automorphism group of $(\infty,2)$-categories. In this note, we will consider the full group $\mathfrak{G} = (\mathbb{Z}_2)^2$ implementing involutory reflections on all levels of morphisms\footnote{Unitary 2-representations of finite groups were already studied extensively in \cite{Bartlett:2008ji}. However, in the language of \cite{Ferrer:2024vpn}, the author of \cite{Bartlett:2008ji} utilises a $\dagger$-structure that corresponds to the choice $\mathfrak{G} = \mathbb{Z}_2$, implementing involutory reflections only on the top level of morphisms. In contrast, in this note we utilise a $\dagger$-structure that corresponds to the choice $\mathfrak{G} = (\mathbb{Z}_2)^2$, implementing involutory reflections on all levels of morphisms. As a result, our construction of unitary 2-representations includes additional coherence data as compared to the one given in \cite{Bartlett:2008ji}.}${}^{,}$\footnote{For 2-categories with all adjoints, there is an enhanced variety of $\dagger$-structures corresponding to different choices of subgroup $\mathfrak{G} \subset \text{Aut}(\text{AdjCat}_{(\infty,2)}) \cong \text{PL}(2)$ \cite{Ferrer:2024vpn}. We will not pursue this direction further.}.
\end{enumerate}

As in the case of local operators, we will without loss of generality restrict attention to unitary 2-representations on finite-dimensional 2-Hilbert spaces, which can be characterised by a finite number $n \in \mathbb{N}$ of simple line operators together with a collection of non-negative Euler terms $\lambda_i \in \mathbb{R}_{> 0}$ ($i=1,...,n$). Since the latter decouple from the action of the global symmetry group $G$, we will henceforth omit them from our discussion and replace the 2-category $\text{2Hilb}$ by the 2-category $\text{Mat}(\text{Hilb})$, whose objects are non-negative integers and morphisms are matrices of Hilbert spaces and linear maps between them. The resulting classification of unitary 2-representations of $G$ can then be summarised as follows:

\textbf{Proposition 1:} The irreducible unitary 2-representations of a finite group $G$ on $\text{Mat}(\text{Hilb})$ can be labelled by triples $\rho = (H,u,q)$ consisting of the following pieces of data:
\begin{enumerate}
\item A subgroup $H \subset G$.
\item A 2-cocycle\footnote{Without loss of generality, we will assume all cochains $c \in C^n(K,U(1))$ to be \textit{normalised} in the sense that $c(k_1,...,k_n)=1$ as soon as $k_i = e$ for at least one $i=1,...,n$.} $u \in Z^2(H,U(1))$.
\item A $H$-covariant\footnote{Here, we define the group of $H$-covariant 1-cochains on $G$ by 
\begin{equation} C^1(G,\mathbb{Z}_2)^H \; := \; \lbrace  q: G \to \mathbb{Z}_2 \; | \; q(h\cdot g) = q(h) \cdot q(g) \;\; \forall \; h \in H, \, g\in G \rbrace \, . 
\end{equation}\vspace{-10pt}} 
1-cochain $q \in C^1(G,\mathbb{Z}_2)^H$.
\end{enumerate}
Two such unitary 2-representations $\rho = (H,u,q)$ and $\rho' = (H',u',q')$ are considered unitarily equivalent if there exists a group element $x \in G$ such that\footnote{We use the notations ${}^gK = gKg^{-1}$ and $K^g = g^{-1}Kg$ for the conjugation of subgroups $K \subset G$ by group elements $g \in G$. Similarly, we write ${}^gh = g h g^{-1}$ and $h^g = g^{-1}\hspace{-1pt}hg$ for $g,h \in G$.}${}^{,}$\footnote{Given $g \in G$ and a cochain $c \in C^n(K,U(1))$ on $K \subset G$, we define the left twist $({}^gc) \in C^n({}^{g\!}K,U(1))$ of $c$ by $g$ by $({}^gc)(k_1,...,k_n) := c(k_1^g,...,k_n^g)$. Similarly, one defines $(c^g) \in C^2(K^g,U(1))$.}
\begin{equation}
H' \, = \, {}^{x\hspace{-1pt}}H \, , \quad\qquad
\left[ \frac{u'}{{}^{x\hspace{-0.3pt}}u} \right] \, = \, 1\, , \quad\qquad q' \, = \, {}^xq \, . 
\end{equation}
The dimension of $\rho = (H,u,q)$ is given by the index $n = |G:H|$ of $H$ in $G$. Upon forgetting the $H$-covariant 1-cochain $q$, this data reduces to the known classification of ordinary 2-representations of $G$ on Kapranov-Voevodsky 2-vector spaces \cite{Kapranov1994,ELGUETA_2007,GANTER20082268,ELGUETA200753,OSORNO2010369}.

From the above, we see that restricting to one-dimensional unitary 2-representations with $H = G$ reproduces the data $(u,q)$ describing the unitary action of $G$ on $G$-invariant line operators as discussed in the previous subsection. Moreover, we can recover the description of twisted sector local operators from the following:

\textbf{Proposition 2:} The irreducible intertwiners between two irreducible unitary 2-represen-tations $\rho = (H,u,q)$ and $\rho' = (H',u',q')$ of $G$ can be labelled by tuples $\eta = (x, \psi)$ consisting of the following pieces of data:
\begin{enumerate}
\item A representative $x \in G$ of a double coset $[x] \in H \backslash G / H'$ such that for all group elements $g \in G$ it holds that
\begin{equation}
q(g) \; = \; \frac{q'(x^{-1}g)}{q'(x^{-1})} \; .
\end{equation}
\item An irreducible unitary representation $\psi$ of $H \cap {}^{x\!}H'$ with projective 2-cocycle
\begin{equation}
\frac{{}^xu'}{u} \;\, \in \,\; Z^2\big(H \cap {}^{x\!}H',U(1)\big) \; .
\end{equation}
\end{enumerate}
In particular, taking $\rho = (G,1,1)$ to be the trivial 2-representation and $\rho'=(G,u',q')$ to be one-dimensional shows that twisted sector local operators at the end of a $G$-invariant line operator transform in unitary projective representations of $G$, provided that the associated reflection anomaly $q'$ vanishes.

The above construction of unitary 2-representations as $\dagger$-2-functors $\rho: BG \to \text{Mat}(\text{Hilb})$ can be generalised in the following ways:
\begin{enumerate}
\item \parbox[t]{\dimexpr\textwidth-\leftmargin}{\vspace{-2.5mm}
\begin{wrapfigure}[7]{r}{0.17\textwidth}
  \vspace{-6pt}
  \begin{center}
    \vspace{-2pt}
    \includegraphics[height=2.9cm]{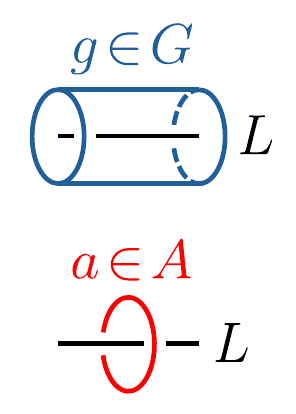}
    \vspace{-2pt}
    \caption{}
    \label{fig-2-group-action}
  \end{center}
\end{wrapfigure}
We can replace the domain of $\rho$ by the delooping of a finite 2-group $\mathcal{G}$. Physically, this corresponds to incorporating an abelian 1-form symmetry $A[1]$ in addition to the 0-form symmetry $G$, where the former acts on line operators via linking (see Figure \ref{fig-2-group-action}).
\vspace{5pt}
\item We can replace the codomain of $\rho$ by the 2-category $\text{Mat}(\text{Herm})$ of matrices of Hermitian spaces\footnotemark. While physically less interesting, this is useful in understanding the role played by positivity in the construction of unitary 2-representations.}
\footnotetext{Here, by \textit{Hermitian space} we mean a finite-dimensional complex vector space $V$ equipped with a non-degenerate sesquilinear form $\braket{.\hspace{1pt}|\hspace{1pt}.}: V \times V \to \mathbb{C}$ satisfying $\braket{v|w} = \braket{w|v}^{\ast}$ for all $v,w \in V$. We denote the category of Hermitian spaces and linear maps between them by $\text{Herm}$ in what follows.}
\end{enumerate}
In this note, we will take both of the above as a starting point by considering $\dagger$-2-functors of the form $\rho: B\mathcal{G} \to \text{Mat}(\text{Herm})$. In order to distinguish these from 2-functors with target $\text{Mat}(\text{Hilb})$, we call the latter \textit{positive unitary} 2-representations, whereas the former are simply called \textit{unitary} 2-representations of $\mathcal{G}$. We provide a full classification of unitary 2-representations and their intertwiners in the main body of this note.

\section{Preliminaries}
\label{sec-preliminaries}

In this section, we review the necessary mathematical ingredients for our discussion of unitary 2-representations. We begin by reviewing the notion of $\dagger$-2-categories in subsection \ref{ssec-dagger-2-categories} and proceed by discussing the two most relevant examples -- the delooping $B\mathcal{G}$ of a finite 2-group $\mathcal{G}$ and the 2-category $\text{2Hilb}$ of 2-Hilbert spaces -- in subsections \ref{ssec-2-groups} and \ref{ssec-2-Hilbert-spaces}.

\subsection[\texorpdfstring{$\dagger$}{x}-2-categories]{\texorpdfstring{$\bm{\dagger}$}{x}-2-categories}
\label{ssec-dagger-2-categories}

The basic ingredients for the constructions described in this note are certain types of 2-categories and 2-functors between them \cite{godement1958,benabou1965,maranda1965}. In general, a 2-category $\mathcal{C}$ consists of a collection of objects $x \in \mathcal{C}$, for each pair of objects $x$ and $y$ a collection of (1-)morphisms $\beta: x \to y$, and for each pair $\beta$ and $\gamma$ of morphisms between objects $x$ and $y$ a collection of 2-morphisms $\Phi: \beta \Rightarrow \gamma$. We often denote this data by 
\begin{equation}
\vspace{-5pt}
\begin{gathered}
\includegraphics[height=1.95cm]{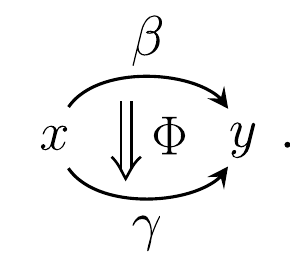}
\end{gathered}
\end{equation}
The (vertical and horizontal) compositions of 1- and 2-morphisms are given by
\begin{equation}
\vspace{-5pt}
\begin{gathered}
\includegraphics[height=4.8cm]{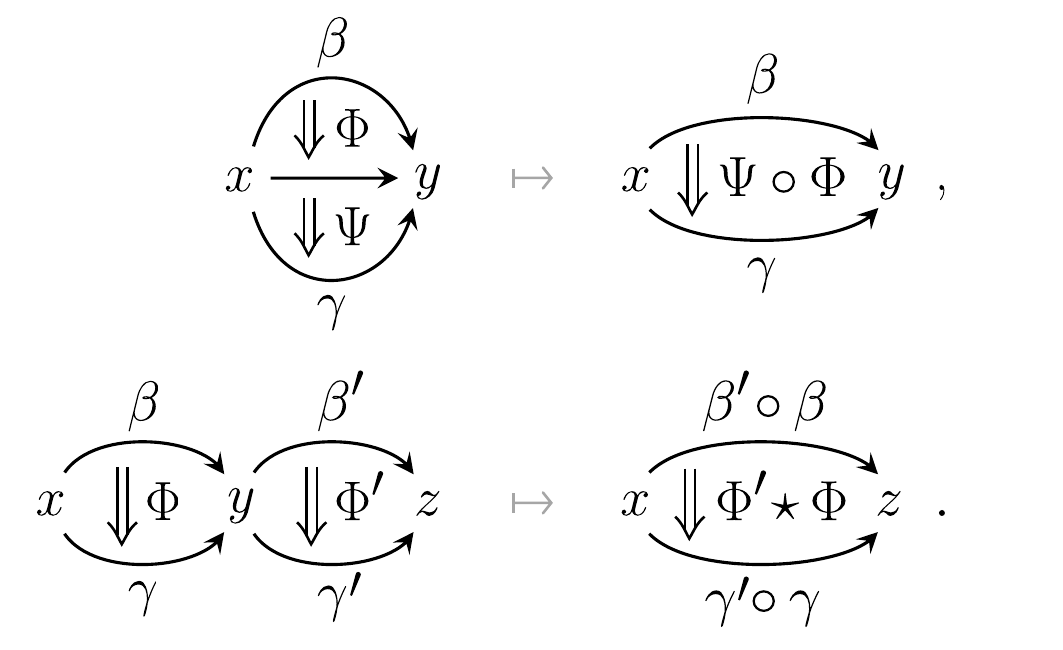}
\end{gathered}
\end{equation}
In this note, we are interested in 2-categories $\mathcal{C}$ that are compatible with reflection positivity, leading to the notion of $\dagger$-2-categories. In general, given an $(\infty,n)$-category $\mathcal{C}$, there is a variety of flavours of higher $\dagger$-structures on $\mathcal{C}$ corresponding to different choices of subgroup $\mathfrak{G} \subset \text{Aut}(\text{Cat}_{(\infty,n)}) = (\mathbb{Z}_2)^n$ \cite{Ferrer:2024vpn}. In this note, we consider the full group $\mathfrak{G} = (\mathbb{Z}_2)^n$, implementing involutory reflections on all levels of morphisms\footnote{For $(\infty,n)$-categories with all adjoints, there is an enhanced variety of $\dagger$-structures corresponding to different choices of subgroup $\mathfrak{G} \subset \text{Aut}(\text{AdjCat}_{(\infty,n)}) \cong \text{PL}(n)$ \cite{Ferrer:2024vpn}. We will not pursue this further.}. In the case $n=2$, this leads to following notion of a $\dagger$-2-category \cite{Ferrer:2024vpn}:

\textbf{Definition:} We call a 2-category $\mathcal{C}$ a \textit{$\dagger$-2-category} if it is equipped with two 2-functors\footnote{Here, $\mathcal{C}^{\text{1op}}$ is the 2-category with the same objects as $\mathcal{C}$ and morphisms $\text{1-Hom}_{\mathcal{C}^{\text{1op}}}(x,y) = \text{1-Hom}_{\mathcal{C}}(y,x)$ for all $x,y \in \mathcal{C}$. Similarly, $\mathcal{C}^{\text{2op}}$ is the 2-category with the same objects and 1-morphisms as $\mathcal{C}$ and 2-morphisms $\text{2-Hom}_{\mathcal{C}^{\text{2op}}}(\beta,\gamma) = \text{2-Hom}_{\mathcal{C}}(\gamma,\beta)$ for all 1-morphisms $\beta$ and $\gamma$ in $\mathcal{C}$.}
\begin{equation}
\dagger_1: \; \mathcal{C} \, \to \, \mathcal{C}^{\text{1op}} \qquad \text{and} \qquad \dagger_2: \; \mathcal{C} \, \to \, \mathcal{C}^{\text{2op}}
\end{equation}
subject to the following conditions:
\begin{itemize}
\item $\dagger_2$ acts as the identity on objects and 1-morphisms and squares to the identity on 2-morphisms, i.e. $(\Phi^{\dagger_2})^{\dagger_2} = \Phi$ for all 2-morphisms $\Phi$ in $\mathcal{C}$.
\item $\dagger_1$ acts as the identity on objects and is equipped with a 2-natural isomorphism $\theta: \dagger_1 \circ \dagger_1 \Rightarrow \text{id}_{\mathcal{C}}$ that is the identity on objects, i.e. only consists of component 2-isomorphisms 
\begin{equation}
\vspace{-5pt}
\begin{gathered}
\includegraphics[height=1.07cm]{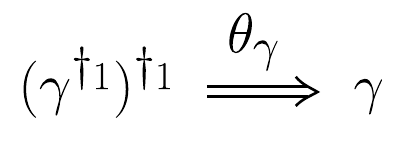}
\end{gathered}
\end{equation}
indexed by 1-morphisms $\gamma$ in $\mathcal{C}$. We require that $(\theta_{\gamma})^{\dagger_1} = \theta_{(\gamma^{\dagger_1})}$ for all $\gamma$.
\end{itemize}
Pictorially, the action of $\dagger_1$ and $\dagger_2$ on morphisms is given by reflections about fixed horizontal and vertical axes, respectively, i.e.
\begin{equation}
\vspace{-5pt}
\begin{gathered}
\includegraphics[height=2.1cm]{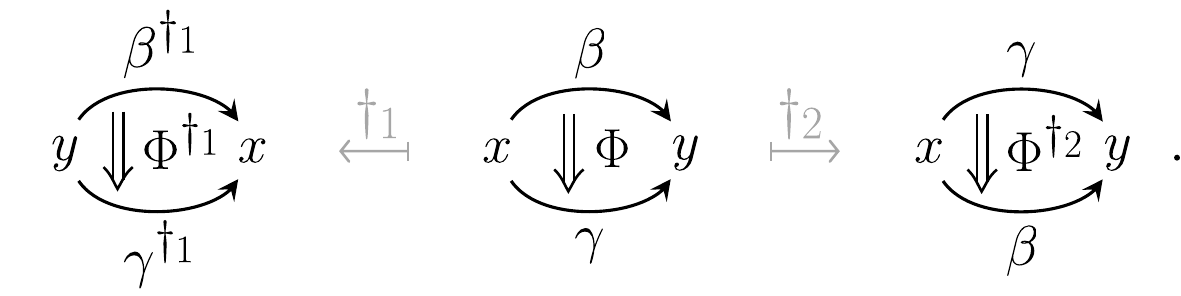}
\end{gathered}
\end{equation}
The above pieces of data need to be compatible with one another in the following sense:
\begin{itemize}
\item $\dagger_1$ and $\dagger_2$ strongly commute, i.e. $(\Phi^{\dagger_1})^{\dagger_2} = (\Phi^{\dagger_2})^{\dagger_1}$ for all 2-morphisms $\Phi$ in $\mathcal{C}$.
\item The component 2-isomorphisms $\theta_{\gamma}$ of the natural transformation $\theta$ are unitary w.r.t. $\dagger_2$ for all 1-morphisms $\gamma$ in $\mathcal{C}$.
\item The compositor 2-isomorphisms $(\dagger_1)_{\beta,\gamma}$ of $\dagger_1$ are unitary w.r.t. $\dagger_2$ for all composable 1-morphisms $\beta$ and $\gamma$ in $\mathcal{C}$.
\item All unitors and associators in $\mathcal{C}$ are unitary w.r.t. $\dagger_2$.
\end{itemize} 
Two objects $x,y \in \mathcal{C}$ are said to be \textit{equivalent} if there exists 1-morphisms $\beta: x \to y$ and $\gamma: y \to x$ such that $\gamma \circ \beta \cong \text{id}_x$ and $\beta \circ \gamma \cong \text{id}_y$. They are said to be \textit{unitarily equivalent} if we can choose $\gamma = \beta^{\dagger_1}$. For a generic 1-morphism $\beta$ in $\mathcal{C}$, we call $\beta^{\dagger_1}$ the \textit{(1-)adjoint} of $\beta$. Similarly, we call $\Phi^{\dagger_1}$ and $\Phi^{\dagger_2}$ the \textit{1-} and \textit{2-adjoint} of a 2-morphism $\Phi$ in $\mathcal{C}$, respectively.

Having introduced a notion of $\dagger$-2-categories\footnote{We note that upon forgetting the 2-functor $\dagger_1$ and its associated coherence data, the definition of a $\dagger$-2-category given above reduces to the notion of a $\dagger$-2-category introduced in \cite{Longo:1996hkk,Chen:2021ttc}. In the language of \cite{Ferrer:2024vpn}, the latter corresponds to a $\dagger$-structure based on the choice of subgroup $\mathfrak{G}=\mathbb{Z}_2$ as opposed to $\mathfrak{G} = (\mathbb{Z}_2)^2$, implementing involutory reflections only on the top level of morphisms.}, we now describe morphisms between them, which correspond to certain types of 2-functors respecting the associated $\dagger$-structures in an appropriate sense. Concretely, we define the following:

\textbf{Definition:} Given two $\dagger$-2-categories $\mathcal{C}$ and $\mathcal{C}'$ as above, a $\dagger$-2-functor between them is a 2-functor\footnote{In this note, we take 2-functors $F: \mathcal{C} \to \mathcal{C}'$ to be \textit{weak} in the sense that they are equipped with natural compositor 2-isomorphisms $F_{\beta,\gamma}: F(\beta) \circ F(\gamma) \Rightarrow F(\beta \circ \gamma)$ for all composable 1-morphisms $\beta$ and $\gamma$.} $F: \mathcal{C} \to \mathcal{C}'$ subject to the following conditions:
\begin{itemize}
\item $F$ commutes with $\dagger_2$, i.e. $F(\Phi^{\dagger_2}) = F(\Phi)^{\dagger'_2}$ for all 2-morphisms $\Phi$ in $\mathcal{C}$.
\item $F$ comes equipped with a 2-natural isomorphism $\jmath: F \circ \dagger_1 \, \Rightarrow \, \dagger_1' \circ F$ that is the identity on objects, i.e. only consists component 2-isomorphisms
\begin{equation}
\vspace{-5pt}
\begin{gathered}
\includegraphics[height=1.07cm]{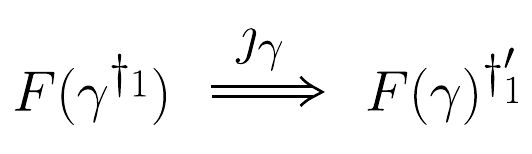}
\end{gathered}
\end{equation}
indexed by 1-morphisms $\gamma$ in $\mathcal{C}$.
\end{itemize}
The above pieces of data need to be compatible with one another in the following ways:
\begin{itemize}
\item The component 2-isomorphisms $\jmath_{\gamma}$ of the natural transformation $\jmath$ are unitary w.r.t. $\dagger'_2$ for all 1-morphisms $\gamma$ in $\mathcal{C}$.
\item The compositor 2-isomorphisms $F_{\beta,\gamma}$ of $F$ are unitary w.r.t. $\dagger_2'$ for all composable 1-morphisms $\beta$ and $\gamma$ in $\mathcal{C}$.
\item $\jmath$ intertwines the 2-natural isomorphisms $\theta$ and $\theta'$ in the sense that the following diagram of 2-natural transformations strictly commutes:
\begin{equation}
\begin{gathered}
\includegraphics[height=2.5cm]{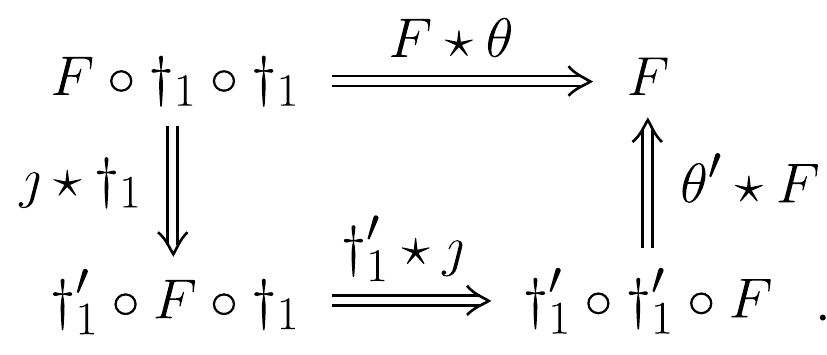}
\end{gathered}
\vspace{4pt}
\end{equation}
\end{itemize}

Given the above notion of $\dagger$-2-functors, we would like to construct a 2-category $[\mathcal{C},\mathcal{C}']^{\dagger}$ of $\dagger$-2-functors between two $\dagger$-2-categories $\mathcal{C}$ and $\mathcal{C}'$. To do this, we need to introduce $\dagger$-2-natural transformations between $\dagger$-2-functors. For the purposes of this note, we achieve this by making the following further assumptions on the target $\dagger$-2-category $\mathcal{C}'$:
\begin{enumerate}
\item We assume that $\mathcal{C}'$ is equipped with \textit{2-duals}, meaning that for each 1-morphism $\gamma: x \to y$ in $\mathcal{C}'$ there exists a 1-morphism $\gamma^{\vee_2}: y \to x$ (the \textit{2-dual} of $\gamma$) together with evaluation and coevaluation 2-morphisms
\begin{equation}
\vspace{-5pt}
\begin{gathered}
\includegraphics[height=1.07cm]{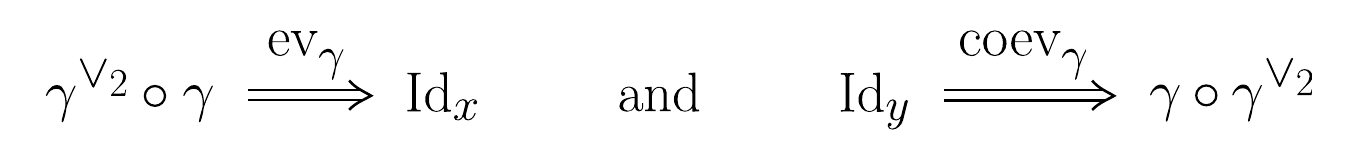}
\end{gathered}
\end{equation}
satisfying suitable zig-zag- or snake relations \cite{Douglas:2018qfz}.
\item We assume that there exists a natural 2-isomorphism $\gamma^{\dagger_1} \cong \gamma^{\vee_2}$ between the 1-adjoint and the 2-dual of each 1-morphism $\gamma$ in $\mathcal{C}'$.
\end{enumerate}
Using these assumptions, we make the following definition: 

\textbf{Definition:} Given two $\dagger$-2-functors $F,\tilde{F}: \mathcal{C} \to \mathcal{C}'$ as above, a \textit{$\dagger$-2-natural transformation} between them is a natural transformation $\eta: F \Rightarrow \tilde{F}$ so that the associated component 1- and 2-morphisms
\begin{equation}
\vspace{-5pt}
\begin{gathered}
\includegraphics[height=2.3cm]{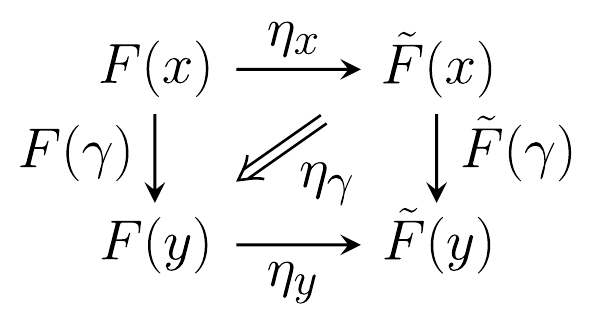}
\end{gathered}
\end{equation}
indexed by objects $x,y \in \mathcal{C}$ and 1-morphisms $\gamma: x \to y$ make the diagram
\begin{equation}
\vspace{-5pt}
\begin{gathered}
\includegraphics[height=6.1cm]{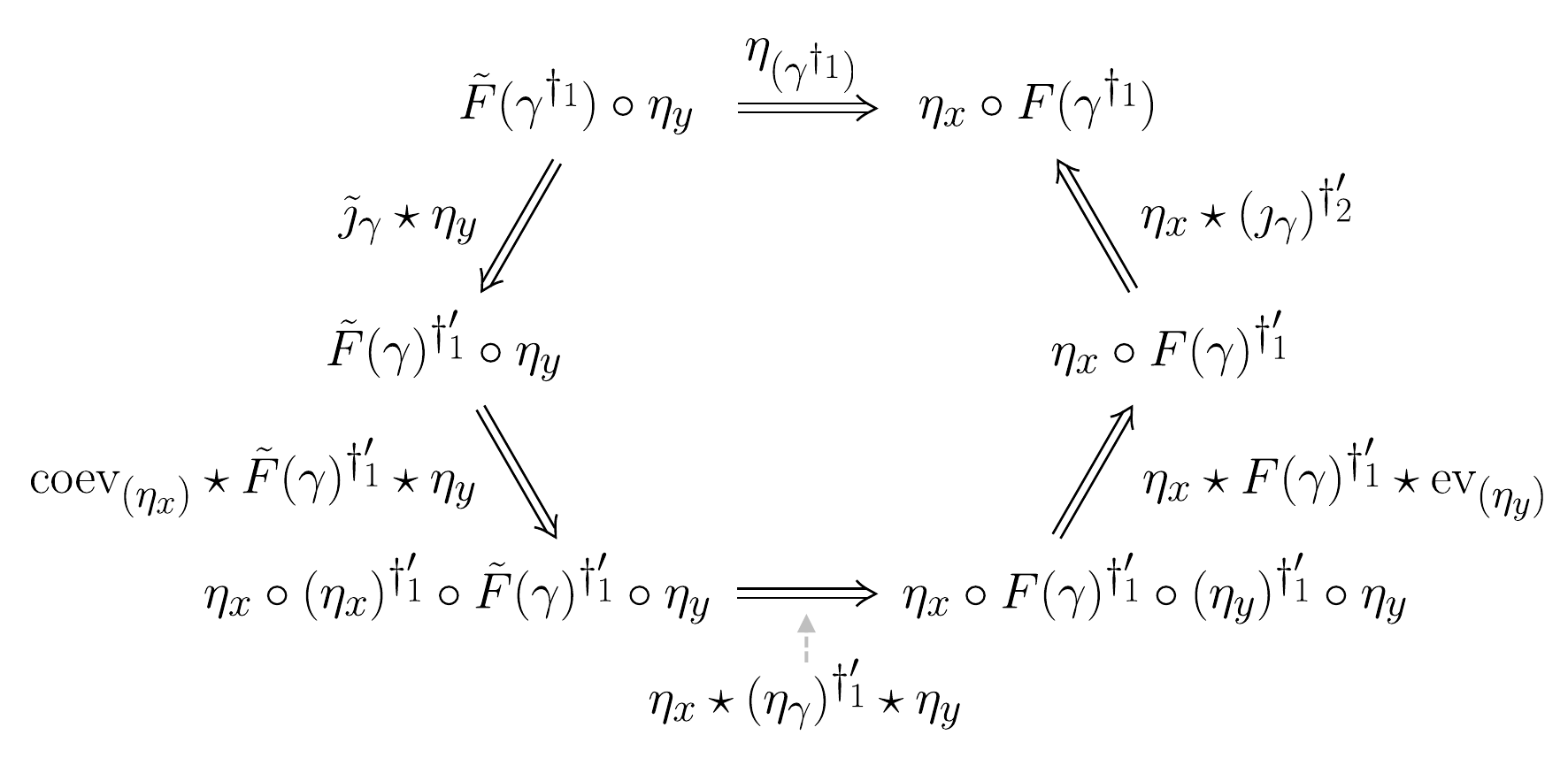}
\end{gathered}
\end{equation}
commute. Here, we implicitly made use of the natural 2-isomorphisms $\gamma^{\dagger_1} \cong \gamma^{\vee_2}$ in $\mathcal{C}'$. 

Using the above, we introduce the 2-category $[\mathcal{C},\mathcal{C}']^{\dagger}$ of $\dagger$-2-functors from $\mathcal{C}$ to $\mathcal{C}'$, their $\dagger$-2-natural transformations and modifications. This 2-category then inherits the structure of a $\dagger$-2-category itself. For instance, given a $\dagger$-2-natural transformation $\eta: F \Rightarrow \tilde{F}$, we construct $\eta^{\dagger_1}: \tilde{F} \Rightarrow F$ to be the $\dagger$-2-natural transformation with component 1-morphisms
\begin{equation}
\vspace{-5pt}
\begin{gathered}
\includegraphics[height=1.07cm]{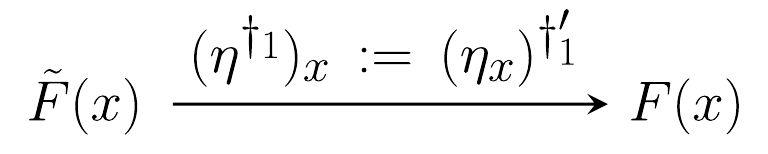}
\end{gathered}
\end{equation}
for every object $x \in \mathcal{C}$ and component 2-morphisms
\begin{equation}
\label{eq-daggered-natural-transformation-component-2}
\vspace{-5pt}
\begin{gathered}
\includegraphics[height=6cm]{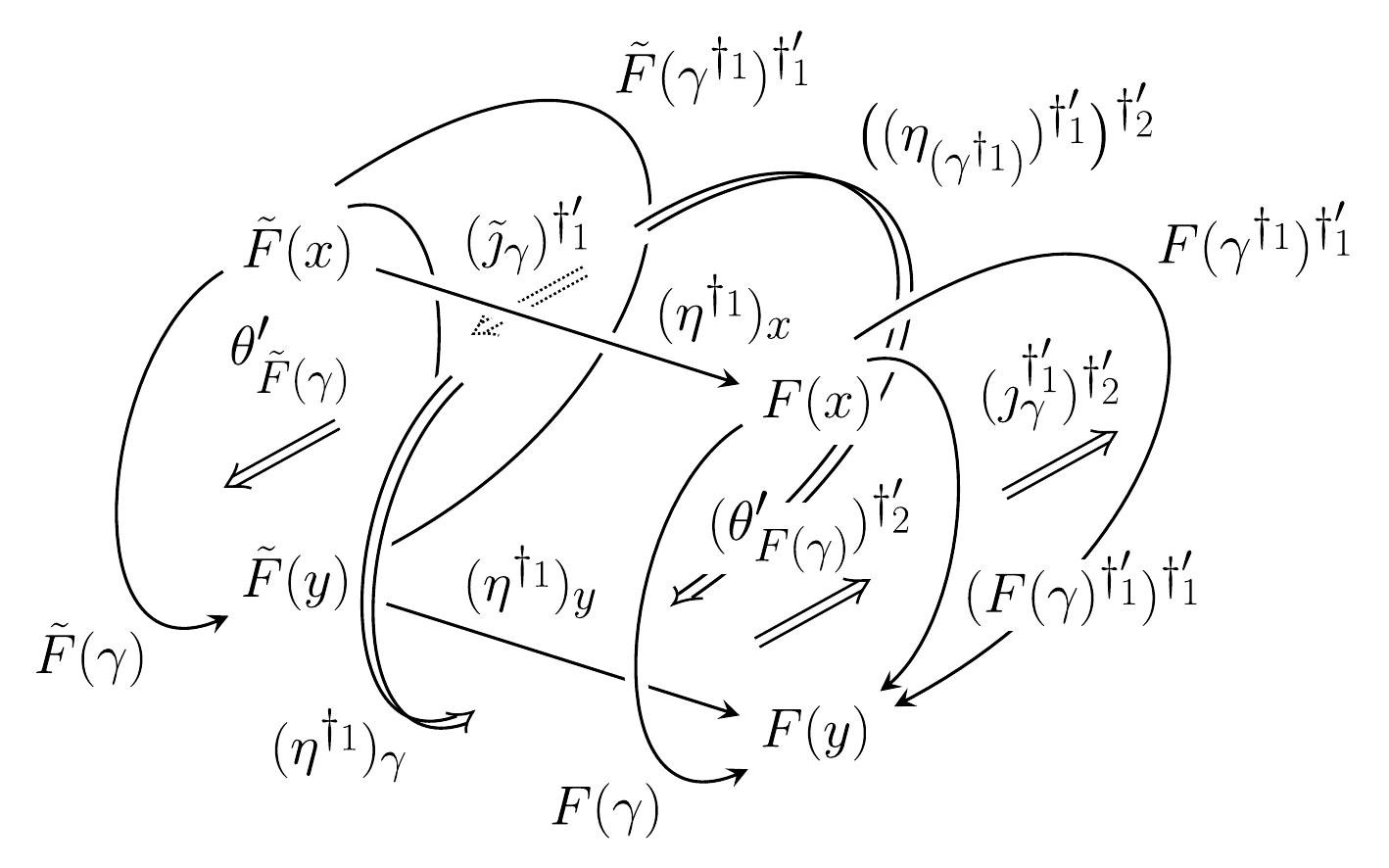}
\end{gathered}
\end{equation}
for every 1-morphism $\gamma: x \to y$ in $\mathcal{C}$. In a similar way, one can construct the action of $\dagger_1$ and $\dagger_2$ on modifications in $[\mathcal{C},\mathcal{C}']^{\dagger}$.

\subsection{2-groups}
\label{ssec-2-groups}

In spacetime dimension $D>2$, line operators can be acted upon by codimension-one (0-form) as well as codimension-two (1-form) symmetry defects \cite{Gaiotto:2014kfa,Delmastro:2022pfo,Bhardwaj:2023wzd,Bartsch:2023pzl,Bartsch2023a,Bhardwaj:2023ayw}. In the finite invertible case, the collection of such defects forms a 2-group $\mathcal{G}$ \cite{Baez2004}, which for the purposes of this note is specified by a quadruple $(G,A,\triangleright,\alpha)$ consisting of
\begin{enumerate}
\item a finite group $G$,
\item a finite abelian group $A$,
\item a group action $\triangleright: G \to \text{Aut}(A)$,
\item a twisted normalised 3-cocycle $\alpha \in Z^3_{\hspace{1pt}\triangleright}(G,A)$.
\end{enumerate}
We will write $\mathcal{G} = A[1] \rtimes_{\alpha} G$ for the 2-group specified by the above data. We further denote by ${}^ga := g \triangleright a$ the group action of an element $g \in G$ on an element $a \in A$. From a physical point of view, $G$ and $A$ represent 0- and 1-form symmetry groups of codimension-one and -two topological defects, respectively, whose interaction is captured by the group action $\triangleright$ and the Postnikov data $\alpha$:
\begin{equation}
\vspace{-5pt}
\begin{gathered}
\includegraphics[height=2.9cm]{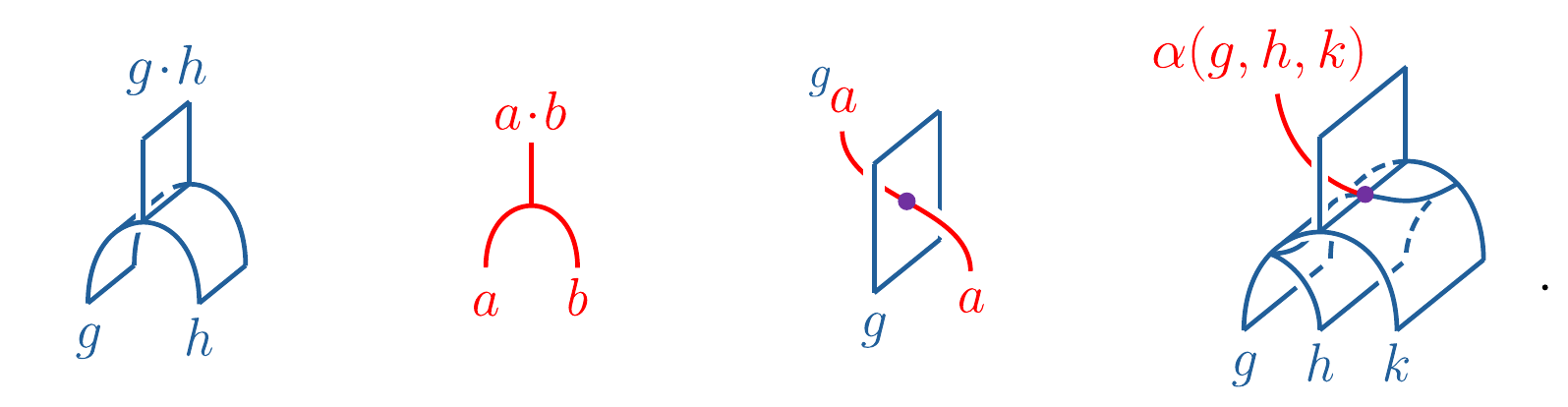}
\end{gathered}
\end{equation}
The existence of 2-group symmetries and their 't Hooft anomalies in quantum field theory has been explored for instance in \cite{Sharpe:2015mja,Kapustin:2013uxa,Cordova:2018cvg,Benini:2018reh,Hsin:2018vcg,Tachikawa:2017gyf,DelZotto:2020esg,Morrison:2020ool,Cordova:2020tij,Bhardwaj:2020phs,Hsin:2020nts,Yu:2020twi,Bhardwaj:2021pfz,Bhardwaj:2021wif,DelZotto:2022joo,Apruzzi:2021vcu,Apruzzi:2021mlh,Brennan:2022tyl,Bhardwaj:2022scy,Bhardwaj:2023zix,Bhardwaj:2022dyt,Liu:2024znj}. To each 2-group $\mathcal{G} = A[1] \rtimes_{\alpha} G$, we can associate a finite monoidal category (which we will also denote by $\mathcal{G}$) that can be described as follows:
\begin{itemize}
\item Its set of objects is given by $G$. The monoidal product $\otimes$ on objects is given by group multiplication in $G$.
\item Its set of morphisms between two objects $g,h \in G$ is given by 
\begin{equation}
\text{Hom}_{\hspace{1pt}\mathcal{G}}(g,h) \; = \; \delta_{g,h} \cdot A
\end{equation}
with composition of morphisms given by group multiplication in $A$. The monoidal product of two morphisms $a \in \text{End}_{\hspace{1pt}\mathcal{G}}(g)$ and $b \in \text{End}_{\hspace{1pt}\mathcal{G}}(h)$ is given by
\begin{equation}
a \otimes b \; = \; a \cdot {}^gb \; .
\end{equation}
\item The associator on three objects $g,h,k \in G$ is given by $\alpha(g,h,k) \in \text{End}_{\hspace{1pt}\mathcal{G}}(g\!\cdot\! h\!\cdot\! k)$.
\end{itemize}
The above monoidal category has the natural structure of an involutive $\dagger$-category\footnote{Following \cite{Egger2011}, an \textit{involutive monoidal category} is a monoidal category $\mathcal{M}$ equipped with an involution functor $\ast: \mathcal{M} \to \mathcal{M}$ and associated natural isomorphisms
\begin{equation}
m^{\ast\ast} \, \cong \, m \quad\qquad \text{and} \quad\qquad n^{\ast} \otimes m^{\ast} \, \cong \, (m \otimes n)^{\ast}
\end{equation} 
for all $m,n \in \mathcal{M}$ satisfying suitable coherence relations. If $\mathcal{M}$ is furthermore a $\dagger$-category, we demand the involution $\ast$ to be compatible with the $\dagger$-structure in the sense that the above isomorphisms are unitary and that $(\omega^{\ast})^{\dagger} = (\omega^{\dagger})^{\ast}$ for all morphisms $\omega$ in $\mathcal{M}$.}, which can be described as follows:
\begin{itemize}
\item On objects $g \in G$ of $\mathcal{G}$, the involution acts as $g^{\ast} := g^{-1}$.
\item On morphisms $a \in \text{End}_{\hspace{1pt}\mathcal{G}}(g) = A$, the involution acts as $a^{\ast} := (a^g)^{-1} \in \text{End}_{\hspace{1pt}\mathcal{G}}(g^{\ast})$. Furthermore, the $\dagger$-structure acts as $a^{\dagger} := a^{-1} \in \text{End}_{\hspace{1pt}\mathcal{G}}(g)$.
\item The involutariness of $\ast$ on objects $g \in G$ of $\mathcal{G}$ is controlled by unitary isomorphisms
\begin{equation}
\label{eq-2-grp-invol-dagger}
\vspace{-5pt}
\begin{gathered}
\includegraphics[height=1.07cm]{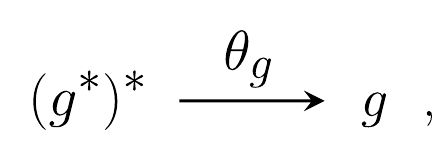}
\end{gathered}
\end{equation}
where we defined the 1-form element $\theta_g(\alpha) := \alpha(g,g^{-1},g)^{-1} \in A$.
\item The compatibility of $\ast$ with the monoidal product of objects $g,h \in G$ of $\mathcal{G}$ is controlled by unitary isomorphisms
\begin{equation}
\label{eq-2-grp-invol-compo}
\vspace{-5pt}
\begin{gathered}
\includegraphics[height=1.07cm]{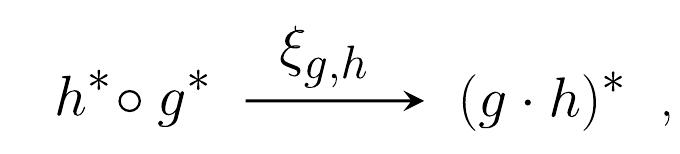}
\end{gathered}
\end{equation}
where we defined the 1-form element
\begin{equation}
\xi_{g,h}(\alpha) \; := \; \frac{\alpha(h^{-1},g^{-1},g) }{\alpha(h^{-1}g^{-1},g,h)} \; \in \; A \; .
\end{equation}
\end{itemize}
Note that the involutive $\dagger$-structure on $\mathcal{G}$ turns its delooping $B\mathcal{G}$ into a $\dagger$-2-category upon identifying $\ast \leftrightarrow \dagger_1$ and $\dagger \leftrightarrow \dagger_2$.

\subsection{2-Hilbert spaces}
\label{ssec-2-Hilbert-spaces}

In contrast to local operators, line operators in a quantum field theory do not form a vector space. While one may define direct sums of line operators using addition of the corresponding correlation functions,
\begin{equation}
\vspace{-5pt}
\begin{gathered}
\includegraphics[height=1.45cm]{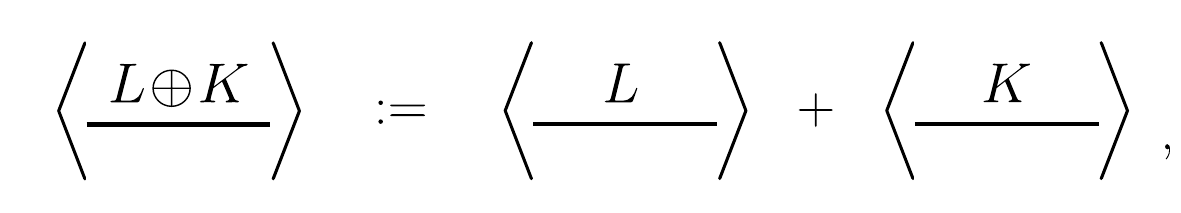}
\end{gathered}
\end{equation}
scalar multiplication of line operators is not well-defined due to their possibly non-trivial internal structure. Concretely, for a given pair of line operators $L$ and $K$, there may exist a non-trivial vector space of topological junction operators 
\begin{equation}
\vspace{-5pt}
\begin{gathered}
\includegraphics[height=1.05cm]{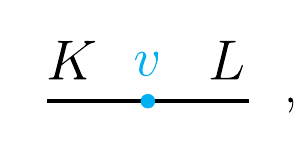}
\end{gathered}
\end{equation}
which we interpret as morphisms between the line $L$ and the line $K$. The collection of line operators together with topological local operators at their junctions then forms a linear abelian category $\mathcal{L}$, whose composition is given by collision of topological junctions, i.e.
\begin{equation}
\vspace{-5pt}
\begin{gathered}
\includegraphics[height=1.45cm]{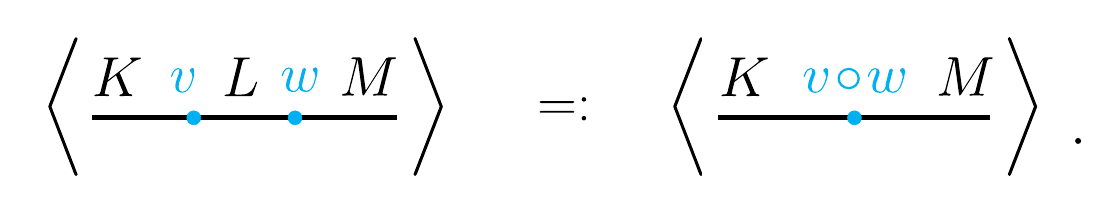}
\end{gathered}
\end{equation}

If the underlying quantum field theory is unitary, the morphism spaces $\text{Hom}_{\mathcal{L}}(L,K)$ inherit additional structure from reflection positivity. Concretely, given a topological junction $v \in \text{Hom}_{\mathcal{L}}(L,K)$, reflecting $v$ about a fixed hyperplane $\Pi$ produces a local operator
\begin{equation}
\vspace{-5pt}
\begin{gathered}
\includegraphics[height=1.67cm]{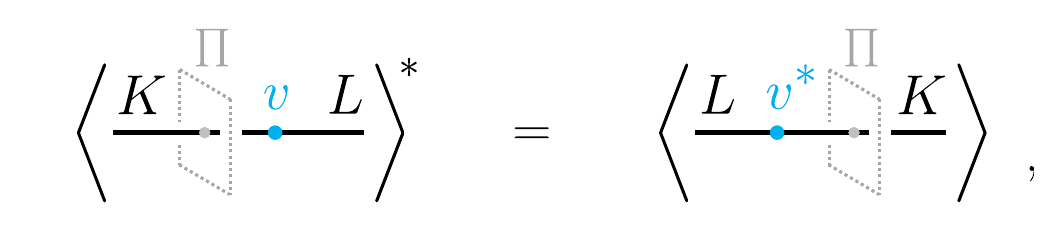}
\end{gathered}
\end{equation}
which induces an antilinear map $\ast: \text{Hom}_{\mathcal{L}}(L,K) \to \text{Hom}_{\mathcal{L}}(K,L)$ satisfying $v^{\ast\ast} = v$ and $(v \circ w)^{\ast} = w^{\ast} \circ v^{\ast}$. Moreover, by performing half-space correlation functions of the type
\begin{equation}
\vspace{-5pt}
\begin{gathered}
\includegraphics[height=1.67cm]{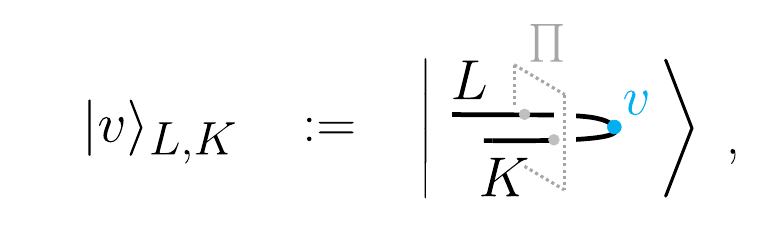}
\end{gathered}
\end{equation}
we obtain a state $\ket{v}_{L,K}$ in the $L$-$K$-twisted Hilbert space, whose inner products can be computed from correlation functions
\begin{equation}
\vspace{-5pt}
\begin{gathered}
\includegraphics[height=1.67cm]{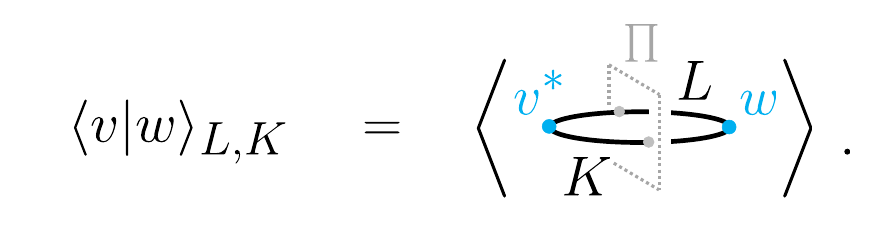}
\end{gathered}
\end{equation}
As a result, the morphism space $\text{Hom}_{\mathcal{L}}(L,K)$ inherits the structure of a Hilbert space, whose inner product $\braket{.\hspace{1pt},.}_{L,K}$ obeys the following relations with the antilinear involution $\ast$:
\begin{equation}
\vspace{-5pt}
\begin{gathered}
\includegraphics[height=4.95cm]{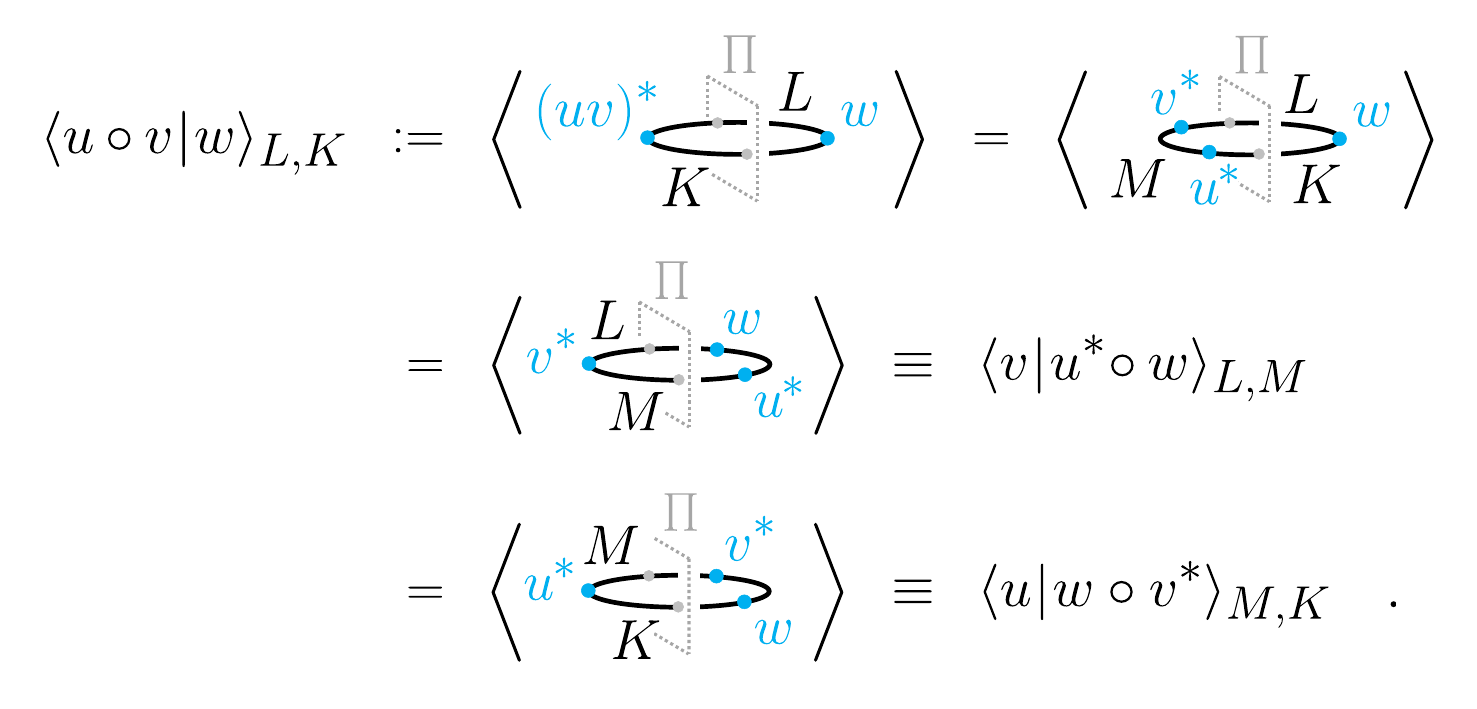}
\end{gathered}
\end{equation}
In analogy to the case of local operators, we call the above structure formed by genuine line operators in a unitary theory a \textit{2-Hilbert space}. This agrees with the mathematical notion of a 2-Hilbert space introduced in \cite{BAEZ1997125} (see also \cite{Freed:1994ad,Bartlett:2008ji,Chen:2024pnk}), which is given by the following:

\textbf{Definition:} A \textit{2-Hilbert space} is an abelian category $\mathcal{L}$ enriched over $\text{Hilb}$ such that for each pair $L,K \in \mathcal{L}$ there exists an antilinear map $\ast: \text{Hom}_{\mathcal{L}}(L,K) \to \text{Hom}_{\mathcal{L}}(K,L)$ satisfying  
\begin{enumerate}
\item $v^{\ast\ast} \, = \, v$,
\item $(v \circ w)^{\ast} \, = \, w^{\ast} \circ v^{\ast}$,
\item $\braket{u\circ v \, | \, w} \, = \, \braket{v \,|\, u^{\ast} \circ w} \, = \, \braket{u \,|\, w \circ v^{\ast}}$
\end{enumerate}
for all morphisms $u,v,w$ in $\mathcal{L}$, whenever both sides of the equation are well-defined. A morphism between 2-Hilbert spaces $\mathcal{L}$ and $\mathcal{L}'$ is a linear functor $F: \mathcal{L} \to \mathcal{L}'$ between the corresponding $\text{Hilb}$-enriched abelian categories such that $F(v^{\ast}) = F(v)^{\ast}$ for all morphisms $v$ in $\mathcal{L}$. A 2-morphism between two 1-morphisms $F$ and $F'$ is a natural transformation. This defines the 2-category $\text{2Hilb}$ of 2-Hilbert spaces\footnote{The notion of a finite-dimensional 2-Hilbert space is intimately related to the notion of a \textit{H*-algebra} \cite{Ambrose1945StructureTF}, which is a Hilbert space $A$ equipped with an associate unital algebra structure and an antilinear involution $\ast: A \to A$ such that $\braket{ab \hspace{1pt}|\hspace{1pt} c}_A = \braket{b\hspace{1pt} | \hspace{1pt}a^{\ast}c}_A = \braket{a \hspace{1pt}|\hspace{1pt} cb^{\ast}}_A$ for all $a,b,c \in A$. Concretely, given a 2-Hilbert space $\mathcal{L}$ with a finite set $\lbrace S_i \rbrace$ of representatives of simple objects, the endomorphism algebra $A := \text{End}_{\mathcal{L}}(\bigoplus_i \hspace{-1pt} S_i)$ is a H*-algebra with antilinear involution given by $\dagger$. Conversely, given a H*-algebra $A$, the category $\text{Mod}^{\dagger}(A)$ of H*-modules over $A$ is naturally a 2-Hilbert space. As a result, we can equivalently view the 2-category of finite-dimensional 2-Hilbert spaces as the 2-category of H*-algebras, their H*-bimodules and bimodule maps.}.

For the purposes of this note, we will restrict attention to finite-dimensional 2-Hilbert spaces corresponding to categories $\mathcal{L}$ with a finite number of simple objects $S_i$ ($i=1,...,n$). The morphism spaces between simple objects are then given by
\begin{equation}
\text{Hom}_{\mathcal{L}}(S_i,S_j) \,\; \cong \,\; \delta_{ij} \cdot \mathbb{C}_{\lambda_i}
\end{equation}
for some $\lambda_i \in \mathbb{R}_{> 0}$, where $\mathbb{C}_{\lambda}$ is isomorphic to $\mathbb{C}$ as an algebra with inner product given by $\braket{v|w} = \lambda \cdot v^{\ast} \cdot w$. Since the parameters $\lambda_i$ decouple from the action of a global symmetry group on simple lines\footnote{A simple line $S$ with a non-trivial parameter $\lambda$ can be viewed as sitting attached to a 2d unitary TQFT with Euler term $\lambda$ \cite{Durhuus:1993cq}. Since we are interested in genuine line operators, we will omit $\lambda$ in what follows.}, we will henceforth omit them from our discussion entirely and regard finite-dimensional 2-Hilbert space $\mathcal{L}$ as being completely determined by its number of simple objects $n \in \mathbb{N}$. Any morphism $F: \mathcal{L} \to \mathcal{L}'$ between 2-Hilbert spaces is then completely determined by its action on simple objects, which is given by
\begin{equation}
F(S_j) \,\; = \,\; \bigoplus_{i\,= \,1}^{n'} \, V_{ij} \otimes S'_i
\end{equation}
for some $(n' \times n)$-matrix $V$ with Hilbert space entries $V_{ij} \in \text{Hilb}$. From a physical perspective, the latter correspond to Hilbert spaces of (non-topological) local operators $\mathcal{O}$ between the simple liner operators $S'_i$ and $S_j$,
\begin{equation}
\vspace{-5pt}
\begin{gathered}
\includegraphics[height=1.05cm]{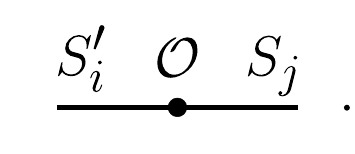}
\end{gathered}
\end{equation}

In summary, for the purposes of this note we can replace the 2-category $\text{2Hilb}$ by the 2-category $\text{Mat}(\text{Hilb})$ of matrices of Hilbert spaces, which can be described as follows:
\begin{itemize}
\item Its objects are non-negative integers $n \in \mathbb{N}$.
\item The (1-)morphisms between objects $m,n \in \mathbb{N}$ are given by $(n \times m)$-matrices $V$ with Hilbert space entries $V_{ij} \in \text{Hilb}$. The composition of two morphisms is given by matrix multiplication using tensor products and direct sums of Hilbert spaces.
\item The 2-morphisms between two 1-morphisms $V$ and $W$ are given by $(n \times m)$-matrices $\Phi$ whose entries are linear maps $\Phi_{ij}: V_{ij} \to W_{ij}$ between the Hilbert space entries of $V$ and $W$. The vertical composition of 2-morphisms $\Phi$ and $\Psi$ is given by entry-wise composition of linear maps. Their horizontal composition is given by matrix multiplication using tensor products and direct sums of linear maps.
\end{itemize}
This 2-category is equipped with 2-duals, where the 2-dual of a 1-morphism $V$ is given by the matrix of Hilbert spaces $V^{\vee_2}$ with entries $(V^{\vee_2})_{ij} = (V_{ji})^{\vee}$, where ${}^{\vee}$ denotes the dual of vector spaces. Furthermore, it naturally possesses the structure of a $\dagger$-2-category:
\begin{itemize}
\item Given a 1-morphism $V: m \to n$, its 1-adjoint is the 1-morphism $V^{\dagger_1}: n \to m$ with entries $(V^{\dagger_1})_{ij} = (V_{ji})^{\ast}$, where ${}^{\ast}$ denotes the complex conjugate of vector spaces. Given a 2-morphism $\Phi: V \Rightarrow W$, its 1-adjoint is the 2-morphism $\Phi^{\dagger_1}: V^{\dagger_1} \Rightarrow W^{\dagger_1}$ with entries $(\Phi^{\dagger_1})_{ij} = (\Phi_{ji})^{\ast}$, where ${}^{\ast}$ denotes the complex conjugate of linear maps.
\item Given a 2-morphism $\Phi: V \Rightarrow W$, its 2-adjoint is the 2-morphism $\Phi^{\dagger_2}: W \Rightarrow V$ with entries $(\Phi^{\dagger_2})_{ij} = (\Phi_{ij})^{\dagger}$, where ${}^{\dagger}$ denotes the adjoint of linear maps.
\end{itemize}
In particular, the Hilbert space structure on the entries of a 1-morphism $V$ in $\text{Mat}(\text{Hilb})$ induces natural isomorphisms $V^{\dagger_1} \cong V^{\vee_2}$ as required. More generally, we can replace the category $\text{Hilb}$ by the categories $\text{Herm}$ or $\text{Vect}$ of Hermitian and ordinary vector spaces\footnote{Replacing $\text{Hilb}$ in $\text{Mat}(\text{Hilb})$ by the category $\text{Vect}$ of finite-dimensional vector spaces recovers (a strictified version of) Kapranov-Voevodsky 2-vector spaces \cite{Kapranov1994,ELGUETA_2007}.}. The canonical functors
$\text{Hilb} \xrightarrow{\,e\,} \text{Herm} \xrightarrow{\,f\,} \text{Vect}$
then induce 2-functors
\begin{equation}
\label{eq-2-hilb-forgetful-2-functors}
\text{Mat}(\text{Hilb}) \;\, \xrightarrow{\;E\;} \,\; \text{Mat}(\text{Herm}) \,\; \xrightarrow{\;F\;} \,\; \text{Mat}(\text{Vect}) \; ,
\end{equation} 
which act as the identity on objects and entry-wise via $e$ and $f$ on 1- and 2-morphisms.

\section{Unitary 2-representations}
\label{sec-unitary-2-representations}

In this section, we discuss the notion of unitary 2-representations of a finite 2-group $\mathcal{G}$ as $\dagger$-2-functors from $B\mathcal{G}$ into certain matrix 2-categories. Concretely, we define the following types of 2-representations of $\mathcal{G}$:
\begin{equation}
\label{eq-def-unitary-2-rep}
\begin{aligned}
\textit{ordinary:} \quad &\text{2Rep}(\mathcal{G}) \; := \; [B\mathcal{G},\text{Mat}(\text{Vect})] \; , \\
\textit{unitary:} \quad &\text{2Rep}^{\dagger}(\mathcal{G}) \; := \; [B\mathcal{G},\text{Mat}(\text{Herm})]^{\dagger} \; , \\
\textit{positive unitary:} \quad &\text{2Rep}^{\dagger}_+(\mathcal{G}) \; := \; [B\mathcal{G},\text{Mat}(\text{Hilb})]^{\dagger} \; .
\end{aligned}
\end{equation}
The 2-functors $E$ and $F$ from (\ref{eq-2-hilb-forgetful-2-functors}) then induce canonical 2-functors
\begin{equation}
\text{2Rep}^{\dagger}_+(\mathcal{G}) \;\, \xrightarrow{\;\;\mathcal{E}\;\;} \,\; \text{2Rep}^{\dagger}(\mathcal{G}) \,\; \xrightarrow{\;\;\mathcal{F}\;\;} \,\; \text{2Rep}(\mathcal{G}) \; .
\end{equation}

We begin this section by discussing the 2-category $\text{2Rep}^{\dagger}(\mathcal{G})$ of unitary 2-representations of $\mathcal{G}$, providing a classification of simple objects and intertwiners between them in subsections \ref{ssec-classification} and \ref{ssec-intertwiners}, respectively. We will discuss the 2-category $\text{2Rep}^{\dagger}_+(\mathcal{G})$ of positive unitary 2-representations of $\mathcal{G}$ in subsection \ref{ssec-positivity}. We conclude by discussing the (positive) unitary 2-representations of the cyclic group $\mathbb{Z}_2$ as an example in subsection \ref{ssec-example}.

\subsection{Classification}
\label{ssec-classification}

In order to classify all unitary 2-representations of a given 2-group $\mathcal{G} = A[1] \rtimes_{\alpha} G$, we list the data associated to a $\dagger$-2-functor $\rho: B\mathcal{G} \to \text{Mat}(\text{Herm})$ below:
\begin{itemize}
\item To the single object $\ast \in B\mathcal{G}$, $\rho$ assigns a non-negative integer $n \in \mathbb{N}$. We will call $n$ the \textit{dimension} of the 2-representation $\rho$ in what follows.
\item To the objects $g \in G$ of $\mathcal{G}$, $\rho$ associates an invertible $(n \! \times \! n)$-matrix of Hermitian spaces $\rho(g)$, which up to equivalence needs to be of the form\footnote{Here, we denote by $\mathbb{C}_{\pm}$ the two simple object of the category $\text{Herm}$ of finite-dimensional complex Hermitian spaces which are isomorphic to $\mathbb{C}$ as a vector space with inner product $\braket{v|w}_{\pm} = \pm \, v^{\ast} \cdot w$.}\textsuperscript{,}\footnote{A priori, the most general form of $\rho(g)$ is $\rho(g)_{ij} = \delta_{\hspace{1pt}i, \hspace{1pt}\sigma_g(j)} \cdot \mathbb{C}_{\tilde{s}_i(g)}$ for some $\tilde{s} \in Z^1_{\sigma}(G,(\mathbb{Z}_2)^n)$. However, up to unitary equivalence, $\tilde{s}$ can always be reabsorbed into the remaining data associated to $\rho$ (see subsection \ref{sssec-equivalences} for a discussion of (unitary) equivalences of unitary 2-representations).}
\begin{equation}
\rho(g)_{ij} \; = \; \delta_{\hspace{1pt}i, \hspace{1pt}\sigma_g(j)} \, \cdot \, \mathbb{C}_+
\end{equation}
for some permutation action $\sigma: G \to S_n$ of $G$ on the finite set $[n]:= \lbrace 1,...,n \rbrace$. We will abbreviate the action of $g \in G$ on indices $i \in [n]$ by $g \triangleright i := \sigma_g(i)$ in what follows.
\item To morphisms $a \in \text{End}_{\mathcal{G}}(g) = A$, $\rho$ assigns an $(n\! \times \!n)$-matrix of unitary linear maps 
\begin{equation}
\vspace{-5pt}
\begin{gathered}
\includegraphics[height=1.07cm]{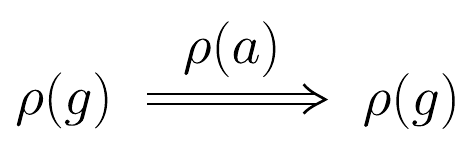}
\end{gathered}
\end{equation}
between the Hermitian space entries of $\rho(g)$, which has to be of the form
\begin{equation}
\rho(a)_{ij} \; = \; \delta_{\hspace{1pt}i, \hspace{1pt}g \hspace{1pt} \triangleright \hspace{1pt} j} \, \cdot \, \chi_i(a)
\end{equation}
for some multiplicative phases $\chi_i(a) \in U(1)$. The latter can be regarded as a collection of characters $\chi \in (A^{\vee})^n$ in the Pontryagin dual group $A^{\vee} := \text{Hom}(A,U(1))$ of $A$, which needs to be compatible with the group action of $G$ on $A$ in the sense that
\begin{equation}
\chi_{g \hspace{1pt} \triangleright \hspace{1pt} i}(a) \; = \; \chi_i(a^g)
\end{equation} 
for all $g \in G$, $a \in A$ and $i \in [n]$.
\item For each pair of objects $g,h \in G$ of $\mathcal{G}$, there exists a unitary 2-isomorphism 
\begin{equation}
\vspace{-5pt}
\begin{gathered}
\includegraphics[height=1.07cm]{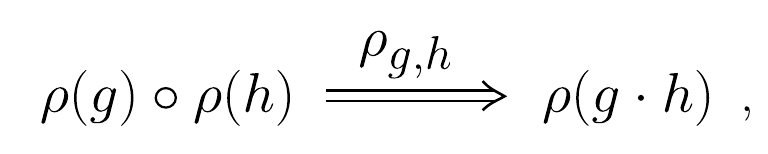}
\end{gathered}
\end{equation}
which needs to be compatible with the monoidal product of three objects $g,h,k \in G$ in the sense that the diagram
\begin{equation}
\label{eq-2-rep-compo-coherence}
\vspace{-5pt}
\begin{gathered}
\includegraphics[height=4.6cm]{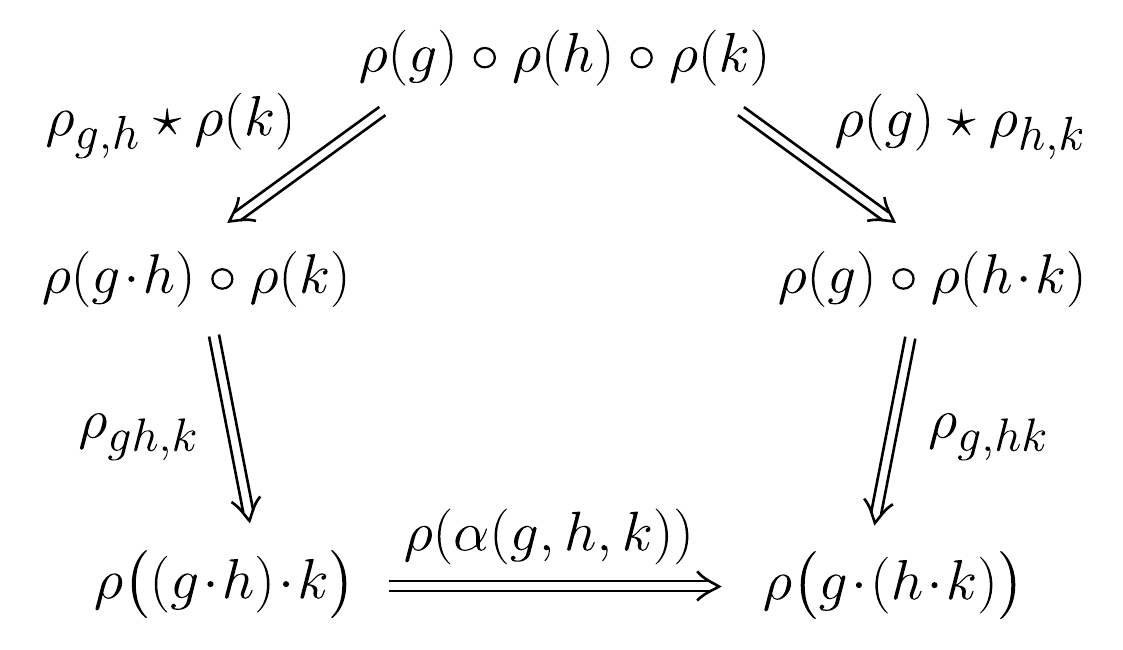}
\end{gathered}
\end{equation}
commutes. Similarly to above, the 2-isomorphisms $\rho_{g,h}$ can then be identified with an invertible $(n\! \times \! n)$-matrix of linear maps of the form
\begin{equation}
(\rho_{g,h})_{ij} \; = \;  \delta_{\hspace{1pt}i, \hspace{1pt}gh \hspace{1pt} \triangleright \hspace{1pt} j} \, \cdot \, c_i(g,h)
\end{equation}
for some multiplicative phases $c_i(g,h) \in U(1)$, which as a consequence of (\ref{eq-2-rep-compo-coherence}) obey
\begin{equation}
\frac{c_{g^{-1} \hspace{1pt} \triangleright \hspace{1pt} i}(h,k) \cdot c_i(g,hk)}{c_i(gh,k) \cdot c_i(g,h)} \; = \; \chi_i(\alpha(g,h,k)) \; .
\end{equation}
The collection of phases $c_i(g,h) \in U(1)$ hence defines a twisted group 2-cochain $c \in C^2_{\sigma}(G,U(1)^n)$ satisfying $d_{\sigma}c = \braket{\chi,\alpha}$. Here, $U(1)^n$ denotes the abelian group that consists of $n$ copies of $U(1)$, acted upon by $G$ via the permutation action $\sigma$. 
\item For each object $g \in G$ of $\mathcal{G}$, there exists a unitary 2-isomorphism
\begin{equation}
\vspace{-5pt}
\begin{gathered}
\includegraphics[height=1.07cm]{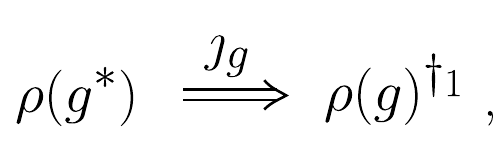}
\end{gathered}
\end{equation}
which needs to be compatible with the $\dagger$-structures on $B\mathcal{G}$ and $\text{Mat}(\text{Herm})$ in the sense that the diagrams
\begin{equation}
\label{eq-2-rep-dagger-coherence}
\vspace{-5pt}
\begin{gathered}
\includegraphics[align=c,height=2.65cm]{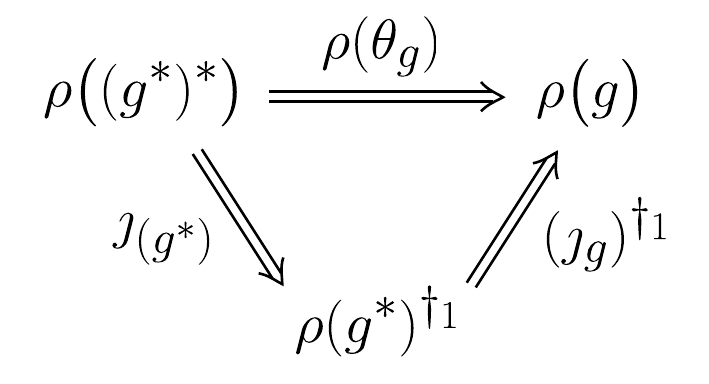}
\qquad
\includegraphics[align=c,height=3.7cm]{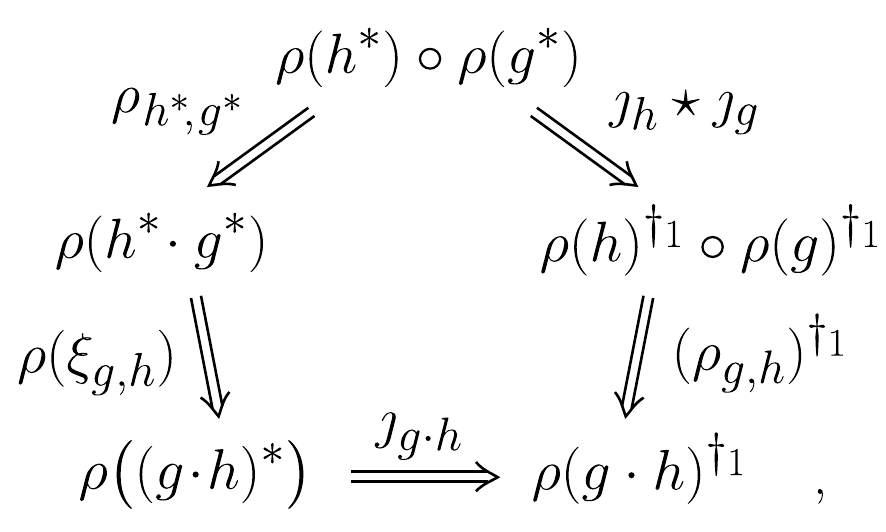}
\end{gathered}
\end{equation}
commute, where $\theta_g$ and $\xi_{g,h}$ are as in (\ref{eq-2-grp-invol-dagger}) and (\ref{eq-2-grp-invol-compo}), respectively. Similarly to above, the 2-isomorphism $\jmath_g$ can then be identified with an invertible $(n\! \times \! n)$-matrix of linear maps of the form 
\begin{equation}
(\jmath_g)_{ij} \; = \; \delta_{\hspace{1pt}g \hspace{1pt} \triangleright \hspace{1pt} i, \hspace{1pt}j} \, \cdot \, \ell_j(g)
\end{equation}
for some multiplicative phases $\ell_j(g) \in U(1)$, which as a consequence of (\ref{eq-2-rep-dagger-coherence}) obey
\begin{gather}
\frac{\ell_{g^{-1} \hspace{1pt} \triangleright \hspace{1pt} i}(g^{-1})}{\ell_i(g)} \;\; = \;\; \chi_i(\theta_g(\alpha)) \; , \label{eq-unitarisation-condition-1} \\
\frac{\ell_{g^{-1}\hspace{1pt}\triangleright \hspace{1pt} i}(h) \cdot \ell_i(g)}{\ell_i(gh)} \;\; = \;\; c_i(g,h) \, \cdot \, c_{(gh)^{-1}\triangleright \hspace{1pt} i}(h^{-1},g^{-1}) \, \cdot \, \chi_{(gh)^{-1}\triangleright \hspace{1pt} i}(\xi_{g,h}(\alpha)) \; . \label{eq-unitarisation-condition-2}
\end{gather}
Note that these conditions can always be solved by
\begin{equation}
(\ell_0)_i(g) \; := \; c_{(g^{-1}) \hspace{0.7pt} \triangleright \hspace{0.7pt} i}(g^{-1},g) \; ,
\end{equation}
with any other solution being of the form
\begin{equation}
\ell \; = \; s \cdot \ell_0
\end{equation}
for some twisted 1-cocycle $s \in Z^1_{\sigma}(G,(\mathbb{Z}_2)^n)$. The space of solutions of conditions (\ref{eq-unitarisation-condition-1}) and (\ref{eq-unitarisation-condition-2}) hence forms a torsor over $Z^1_{\sigma}(G,(\mathbb{Z}_2)^n)$.
\end{itemize}

To summarise, a unitary 2-representation of $\mathcal{G} = A[1] \rtimes_{\alpha} G$ can be labelled by quintuples consisting of the following pieces of data:
\begin{enumerate}
\item A non-negative integer $n \in \mathbb{N}$, called the \textit{dimension} of the 2-representation.
\item A permutation action $\sigma: G \to S_n$ of $G$ on $[n] := \lbrace 1,...,n \rbrace$.
\item A collection of $n$ characters $\chi \in (A^{\vee})^n$ satisfying $\chi_{g \hspace{1pt} \triangleright \hspace{1pt} i}(a) = \chi_i(a^g)$.
\item A twisted 2-cochain $c \in C^2_{\sigma}(G,U(1)^n)$ satisfying $d_{\sigma}c = \braket{\chi,\alpha} $.
\item A twisted 1-cocycle $s \in Z^1_{\sigma}(G,(\mathbb{Z}_2)^n)$.
\end{enumerate}
We will write $\rho = (n,\sigma,\chi,c,s)$ for a unitary 2-representation specified by the above data in what follows. The \textit{dual} of $\rho$ is the unitary projective 2-representation $\rho^{\vee_1}$ specified by
\begin{equation}\label{eq-dual-2-representation}
\rho^{\vee_1} \; = \; (n,\sigma, \chi^{\ast},c^{\ast},s) \; ,
\end{equation}  
where ${}^{\ast}$ denotes complex conjugation. The trivial 2-representation of $\mathcal{G}$ is the unitary 2-representation with associated data $\one = (1,1,1,1,1)$. 

In terms of the above classification, the canonical 2-functor $\mathcal{F}\!: \text{2Rep}^{\dagger}(\mathcal{G}) \to \text{2Rep}(\mathcal{G})$ sends
\begin{equation}
(n,\sigma,\chi,c,s) \; \mapsto \; (n,\sigma,\chi,c) \; ,
\end{equation}
which reproduces the known classification of ordinary 2-representations of $\mathcal{G}$ on on Kapra-nov-Voevodsky 2-vector spaces by quadruples $(n,\sigma,\chi,c)$ \cite{GANTER20082268,ELGUETA200753,OSORNO2010369}. In particular, $\mathcal{F}$ is essentially surjective, which can be seen as a higher analogue of the fact that any finite-dimensional representation of a finite group $G$ is equivalent to a unitary one.

\subsubsection{Irreducibles}
\label{sssec-irreducible-2-reps}

A unitary 2-representation $\rho = (n,\sigma,\chi,c,s)$ of $\mathcal{G}$ is irreducible if the associated permutation action $\sigma: G \to S_n$ is transitive. In this case, we can use the orbit-stabiliser theorem to relate the $G$-orbit $[n] \equiv \lbrace 1,...,n\rbrace$ to the stabiliser subgroup
\begin{equation}
H \; := \; \text{Stab}_{\sigma}(1) \; \equiv \; \lbrace h \in G \; | \; \sigma_h(1) = 1 \rbrace \; \subset \; G
\end{equation}
of a fixed element $1 \in [n]$. The remaining data associated to $\rho$ then gives rise to a one-dimensional unitary 2-representation of the sub-2-group $\mathcal{H} = A[1] \rtimes_{(\alpha|_H)} H \subset \mathcal{G}\,$:
\begin{itemize}
\item By setting $\lambda:= \chi_1$, we obtain a character $\lambda \in A^{\vee}$ that is $H$-invariant in the sense that $\lambda({}^ha) = \lambda(a)$ for all $h \in H$ and $a \in A$.
\item By setting $u := c_1|_H$, we obtain a 2-cochain $u \in C^2(H,U(1))$ obeying $du \; = \; \braket{\lambda,\alpha|_H}$.
\item By setting $p := s_1|_H$, we obtain a homomorphism $p \in Z^1(H,\mathbb{Z}_2) \equiv \text{Hom}(H,\mathbb{Z}_2)$.
\end{itemize}

Conversely, given a subgroup $H \subset G$ and a one-dimensional unitary 2-representation $(\lambda,u,p)$ of the sub-2-group $\mathcal{H} = A[1] \rtimes_{(\alpha|_H)} H$, we can construct an irreducible unitary 2-representation $\rho = (n,\sigma,\chi,c,s)$ of $\mathcal{G} = A[1] \rtimes_{\alpha} G$ via induction; 
\begin{equation}
(n,\sigma,\chi,c,s) \; = \; \text{Ind}_{\hspace{1pt}\mathcal{H}}^{\hspace{1pt}\mathcal{G}}(\lambda,u,p) \; .
\end{equation}
To this end, let $\lbrace r_1,...,r_n \rbrace$ be a fixed set of representatives $r_i \in G$ of left cosets of $H$ in $G$,
\begin{equation}
G/H \; = \; \lbrace r_1 H, ..., r_n H \rbrace \, ,
\end{equation}
so that $r_1 = e$ is the identity element and $n = |G:H|$ is the index of $H$ in $G$. From this, we can obtain the data of an irreducible unitary 2-representation of $\mathcal{G}$ as follows:
\begin{itemize}
\item Given the set of fixed representatives $r_i$ of left $H$-costes in $G$, left multiplication by group elements $g \in G$ induces a permutation action $\sigma: G \to S_n$ via
\begin{equation}
g \cdot r_iH \; = \; r_{\sigma_g(i)}H \, ,
\end{equation}
which we abbreviate by $g \triangleright i := \sigma_g(i)$ in what follows. This then allows us to define for each $g \in G$ and $i \in [n]$ an associated little group element
\begin{equation}
g_i \; := \; r_i^{-1} \cdot g \cdot r_{(g^{-1}) \hspace{0.7pt} \triangleright \hspace{0.7pt} i} \; \in \; H \; .
\end{equation}
\item Given the $H$-invariant character $\lambda \in A^{\vee}$, we obtain a collection $\chi \in (A^{\vee})^n$ of characters via $\chi_i(a) :=  \lambda(a^{r_i})$ satisfying $\chi_{g \hspace{1pt} \triangleright \hspace{1pt} i}(a) = \chi_i(a^g)$.
\item Given the 2-cochain $u \in C^2(H,U(1))$ obeying $du \; = \; \braket{\lambda,\alpha|_H}$, we obtain a twisted 2-cochain $c \in C^2_{\sigma}(G,U(1)^n)$ obeying $d_{\sigma}c = \braket{\chi,\alpha}$ by setting
\begin{equation}\label{eq-induced-2-rep-2-cocycle}
c_i(g,h) \; := \; \big\langle\lambda, \phi_i(\alpha)(g,h)\big\rangle \, \cdot \, u\big( g_i , h_{g^{-1} \hspace{0.7pt}\triangleright \hspace{0.7pt} i} \big) \; ,
\end{equation}
where we defined the multiplicative factor
\begin{equation}
\phi_i(\alpha)(g,h) \; := \; \frac{\alpha(r_i^{-1},g,h) \, \cdot \, \alpha\big(g_i, \, h_{g^{-1} \hspace{0.7pt}\triangleright \hspace{0.7pt} i} , \, r^{-1}_{(gh)^{-1} \hspace{0.7pt}\triangleright \hspace{0.7pt} i} \big)}{\alpha\big( g_i , \, r^{-1}_{g^{-1} \hspace{0.7pt}\triangleright \hspace{0.7pt} i} , \, h \big)} \; \in \; A \; .
\end{equation}
\item Given the group homomorphism $p \in \text{Hom}(H,\mathbb{Z}_2)$, we obtain a twisted 1-cocycle $s \in Z^1_{\sigma}(G,(\mathbb{Z}_2)^n)$ by setting $s_i(g) := p(g_i)$.
\end{itemize}
To summarise, we can label the irreducible unitary 2-representations of $\mathcal{G} = A[1] \rtimes_{\alpha} G$ by quadruples consisting of the following pieces of data:
\begin{enumerate}
\item A subgroup $H \subset G$.
\item A $H$-invariant character $\lambda \in A^{\vee}$.
\item A 2-cochain $u \in C^2(H,U(1))$ satisfying $du = \braket{\lambda,\alpha|_H}$.
\item A group homomorphism $p \in \text{Hom}(H,\mathbb{Z}_2)$.
\end{enumerate}
We will write $\rho = (H,\lambda,u,p)$ for an irreducible unitary 2-representation of $\mathcal{G}$ specified by the above data in what follows. The dimension of such a 2-representation is given by the index $n = |G:H|$ of $H$ in $G$. The \textit{dual} of $\rho$ is given by
\begin{equation}
\rho^{\vee_1} \; = \; (H,\lambda^{\ast},u^{\ast},p) \; ,
\end{equation}
where ${}^{\ast}$ denotes complex conjugation. The trivial 2-representation of $\mathcal{G}$ has associated data given by $\one = (G,1,1,1)$. The canonical 2-functor $\mathcal{F}\!: \text{2Rep}^{\dagger}(\mathcal{G}) \to \text{2Rep}(\mathcal{G})$ sends
\begin{equation}
(H,\lambda,u,p) \; \mapsto \; (H,\lambda,u) \; ,
\end{equation}
which reproduces the known classification of ordinary irreducible 2-representations of $\mathcal{G}$ on Kapranov-Voevodsky 2-vector spaces by triples $(H,\lambda,u)$ \cite{GANTER20082268,ELGUETA200753,OSORNO2010369} (see also \cite{Bartsch:2022mpm,Bartsch:2022ytj,Bhardwaj:2022lsg,Bhardwaj:2022kot,Bhardwaj:2022maz} for a physical interpretation of 2-representations as Wilson surfaces in the context of the discrete gauging of finite invertible symmetries in three dimensions).

\subsection{Intertwiners}
\label{ssec-intertwiners}

In order to discuss equivalences of unitary 2-representations, we need to introduce the notion of intertwiners between them. Using (\ref{eq-def-unitary-2-rep}), an intertwiner between two unitary 2-representations $\rho = (n,\sigma,\chi,c,s)$ and $\rho' = (n',\sigma',\chi',c',s')$ is given by a $\dagger$-2-natural transformation $\eta: \rho \Rightarrow \rho'$, whose associated data can be described as follows:
\begin{itemize}
\item To the single object $\ast \in B\mathcal{G}$, $\eta$ assigns a morphism $\eta_{\ast}$ between $\rho(\ast)= n$ and $\rho'(\ast) = n'$, which can be identified with an $(n'\! \times\! n)$-matrix $V$ with Hermitian space entries $V_{ij}$.
\item To the objects $g \in G$ of $\mathcal{G}$, $\eta$ assigns 2-morphisms
\begin{equation}
\vspace{-5pt}
\begin{gathered}
\includegraphics[height=1.07cm]{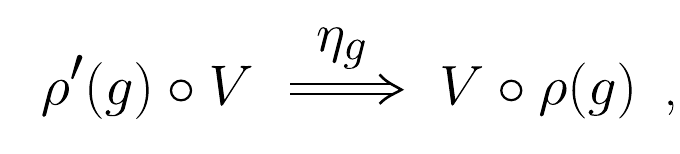}
\end{gathered}
\end{equation}
which need to be compatible with the monoidal product of two objects $g,h \in G$ in the sense that the diagram
\begin{equation}
\label{eq-intertwiner-compo-coherence}
\vspace{-5pt}
\begin{gathered}
\includegraphics[height=4.15cm]{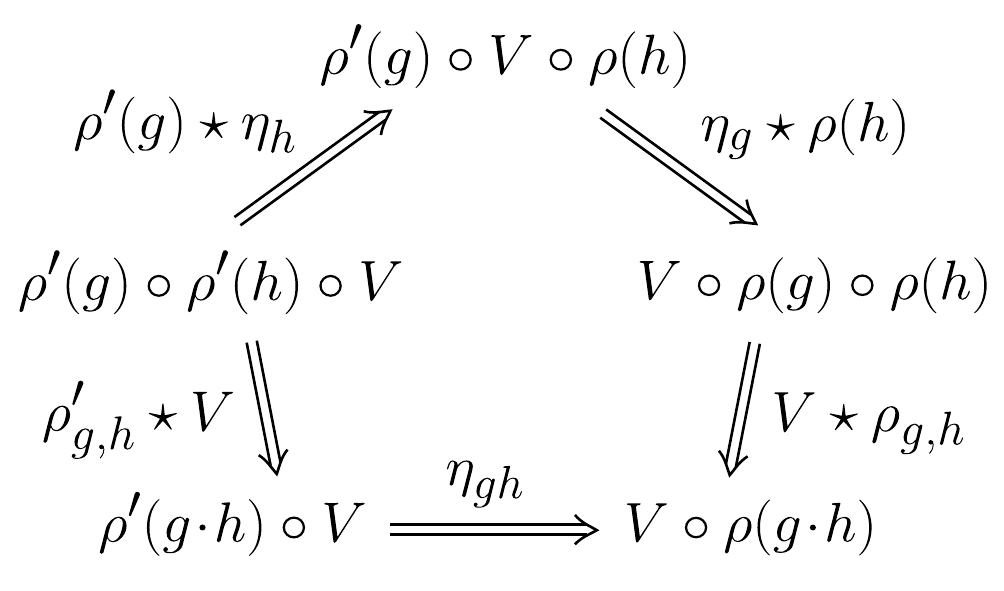}
\end{gathered}
\end{equation}
commutes. Upon identifying $\eta_g$ with an $(n' \! \times \! n)$-matrix of linear maps
\begin{equation}
(\eta_g)_{ij} \; =: \; \varphi(g)_{(\sigma'_{g^{-1}})(i),\, j}
\end{equation}
with $\varphi(g)_{ij}: V_{ij} \to V_{g \, \triangleright \, (i,j)}$, condition (\ref{eq-intertwiner-compo-coherence}) becomes equivalent to
\begin{equation}
\varphi(g)_{h \, \triangleright (i,j)} \ \circ \, \varphi(h)_{ij} \,\; = \,\; \frac{c'_{gh \, \triangleright i}(g,h)}{c_{gh \, \triangleright j}(g,h)} \, \cdot \, \varphi(g \cdot h)_{ij} \; ,
\end{equation}
where we denoted by $g \,\triangleright (i,j) := (\sigma'_g(i), \sigma_g(j))$ the product action $\sigma' \times \sigma$ of $G$ on $[n'] \times [n]$. Furthermore, in order for $\eta$ to be compatible with the action of the 1-form symmetry group $A$, the diagram
\begin{equation}
\vspace{-5pt}
\begin{gathered}
\includegraphics[height=2.5cm]{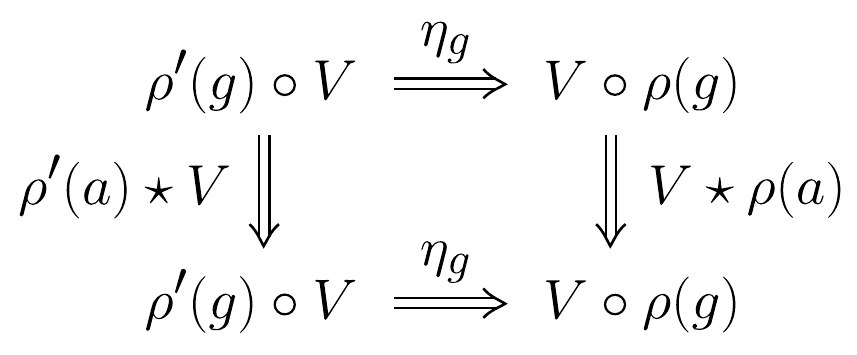}
\end{gathered}
\end{equation}
has to commute for all $a \in \text{End}_{\hspace{1pt}\mathcal{G}}(g) = A$, leading to the condition
\begin{equation}
\chi'_{g\, \triangleright i}(a) \,\cdot\, \varphi(g)_{ij} \; = \; \chi_{g \, \triangleright j}(a) \,\cdot\, \varphi(g)_{ij} \; .
\end{equation}
In particular, setting $g=e$ (so that $\varphi(e)_{ij} = \text{id}_{V_{ij}}$) reveals that $V_{ij}=0$ unless $\chi'_i = \chi_j \in A^{\vee}$. Lastly, for $\eta$ to be a $\dagger$-2-natural transformation, we require it to be compatible with the involution on objects $g \in G$ of $\mathcal{G}$ in the sense that the diagram
\begin{equation}
\vspace{-5pt}
\begin{gathered}
\includegraphics[height=6.1cm]{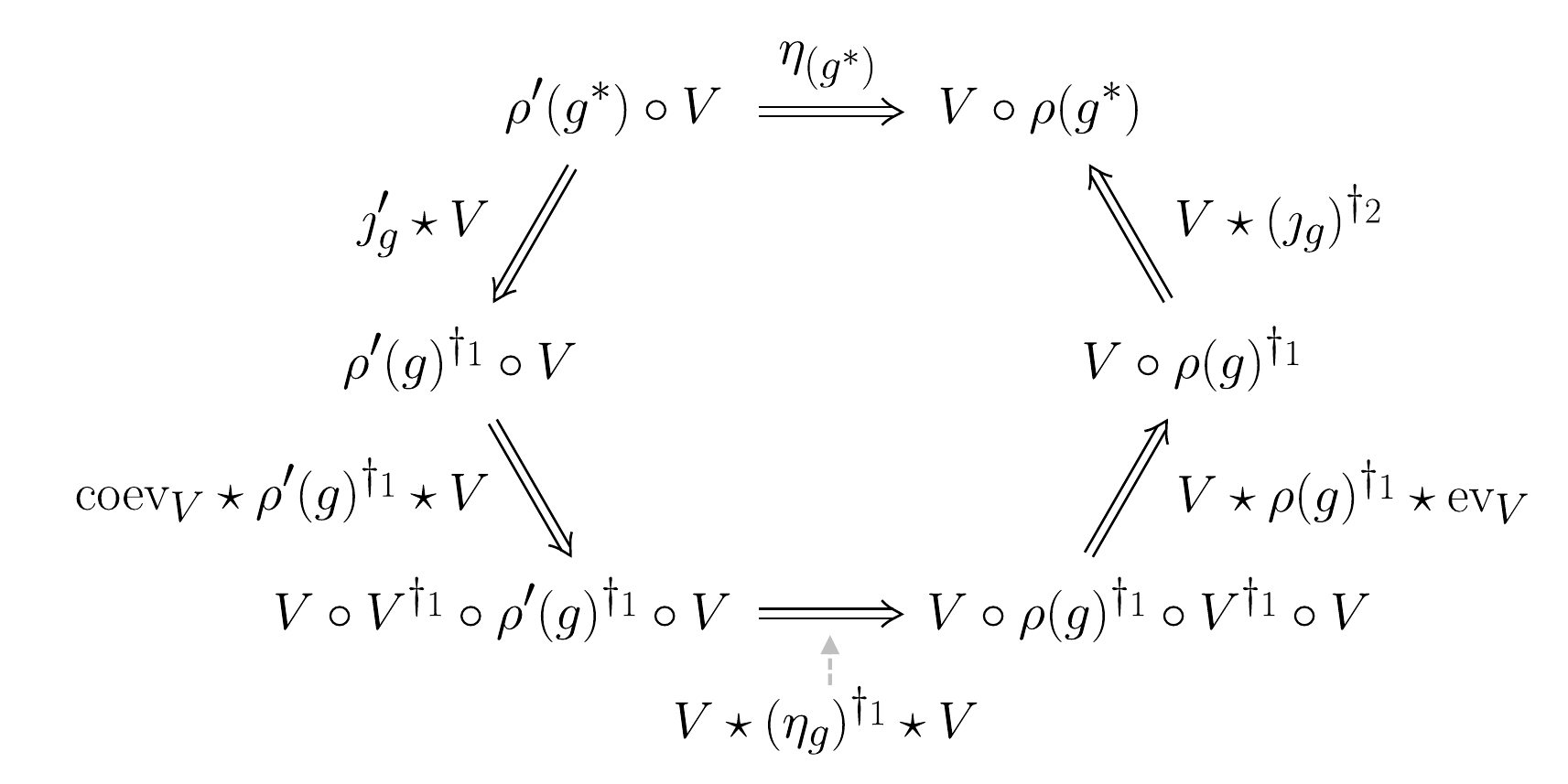}
\end{gathered}
\end{equation}
commutes, where we implicitly made use of the natural isomorphism $V^{\dagger_1} \cong V^{\vee_2}$ in $\text{Mat}(\text{Herm})$. This then translates to the condition
\begin{equation}
\varphi(g)_{ij}^{\dagger} \; = \; \frac{s_{g\, \triangleright j}(g)}{s'_{g\, \triangleright i}(g)} \, \cdot \, \varphi(g)_{ij}^{-1} \; ,
\end{equation}
where ${}^{\dagger}$ denotes the adjoint of linear maps between Hermitian spaces.
\end{itemize}
To summarise, intertwiners $\eta$ between two unitary 2-representations $\rho = (n,\sigma,\chi,c,s)$ and $\rho' = (n',\sigma',\chi',c',s')$ can be labelled by tuples consisting of the following pieces of data:
\begin{enumerate}
\item An $(n'\! \times \! n)$-matrix of Hermitian spaces $V_{ij}$ with $V_{ij} = 0$ unless $\chi_i' = \chi_j$.
\item For each $g \in G$ a collection of linear maps $\varphi(g)_{ij}: V_{ij} \to V_{g \, \triangleright (i,j)}$ that satisfy the composition rule
\begin{equation}\label{eq-intertwiner-composition-rule}
\varphi(g)_{h \, \triangleright \hspace{1pt} (ij)} \ \circ \, \varphi(h)_{ij} \,\; = \,\; \frac{c'_{gh \, \triangleright i}(g,h)}{c_{gh \, \triangleright j}(g,h)} \, \cdot \, \varphi(g \cdot h)_{ij}
\end{equation}
as well as the conjugation rule
\begin{equation}\label{eq-intertwiner-conjugation-rule}
\varphi(g)_{ij}^{\dagger} \; = \; \frac{s_{g\, \triangleright j}(g)}{s'_{g\, \triangleright i}(g)} \, \cdot \, \varphi(g)_{ij}^{-1} \; .
\end{equation}
\end{enumerate}
We will write $\eta = (V,\varphi)$ for an intertwiner specified by the above data in what follows. The identity morphism of a unitary 2-representation $\rho$ is the intertwiner $\text{id}_{\rho}: \rho \Rightarrow \rho$ with associated data $\text{id}_{\rho} = (\one_n, \text{Id}_n)$, where
\begin{equation}
(\one_n)_{ij} \; = \; \delta_{ij} \cdot \mathbb{C} \qquad \text{and} \qquad \text{Id}_n(g)_{ij} \; = \; \delta_{ij} \cdot \text{id}_{\mathbb{C}} \; .
\end{equation}
The duals and adjoints of an intertwiner $\eta = (V,\varphi): \rho \Rightarrow \rho'$ can be described as follows:
\begin{itemize}
\item The \textit{1-dual} of $\eta$ is defined to be the intertwiner $\eta^{\vee_1}: (\rho')^{\vee_1} \Rightarrow \rho^{\vee_1}$ that has associated data $\eta^{\vee_1} = (V^{\vee_1}, \varphi^{\vee_1})$ with $(V^{\vee_1})_{ij}  = V_{ji}$ and
\begin{equation}
\label{eq-intertwiner-1-dual}
(\varphi^{\vee_1})(g)_{ij} \; = \; \varphi(g)_{ji} \; .
\end{equation}
\item The \textit{2-dual} of $\eta$ is defined to be the intertwiner $\eta^{\vee_2}: \rho' \Rightarrow \rho$ that has associated data $\eta^{\vee_2} = (V^{\vee_2}, \varphi^{\vee_2})$ with $(V^{\vee_2})_{ij} = (V_{ji})^{\vee}$ and
\begin{equation}
\label{eq-intertwiner-2-dual}
(\varphi^{\vee_2})(g)_{ij} \; = \; (\varphi(g)_{ji}^{-1})^{\vee} \; ,
\end{equation}
where ${}^{\vee}$ denotes the transpose of linear maps between vector spaces.
\item Using (\ref{eq-daggered-natural-transformation-component-2}), the \textit{adjoint} of $\eta$ can be computed to be the intertwiner $\eta^{\dagger_1}: \rho' \Rightarrow \rho$ that is specified by data $\eta^{\dagger_1} = (V^{\dagger_1},\varphi^{\dagger_1})$ with $(V^{\dagger_1})_{ij} = (V_{ji})^{\ast}$ and
\begin{equation}\label{eq-intertwiner-adjoint}
(\varphi^{\dagger_1})(g)_{ij} \;\, = \;\, \varphi(g)_{ji}^{\ast} \; ,
\end{equation}
where ${}^{\ast}$ denotes the complex conjugate of linear maps between vector spaces.
\end{itemize}

\subsubsection{Irreducibles}
\label{sssec-irreducible-intertwiners}

In order to classify the irreducible intertwiners between two irreducible unitary 2-represen-tations $\rho = (n,\sigma,\chi,c,s)$ and $\rho' = (n',\sigma',\chi',c',s')$ of $\mathcal{G}$, we write the latter as inductions
\begin{equation}
\rho \; = \; \text{Ind}_{\mathcal{H}}^{\mathcal{G}}(\lambda,u,p) \qquad \text{and} \qquad \rho' \; = \; \text{Ind}_{\mathcal{H}'}^{\mathcal{G}}(\lambda',u',p')
\end{equation}
of one-dimensional unitary 2-representations $(\lambda,u,p)$ and $(\lambda',u',p')$ of certain sub-2-groups $\mathcal{H}^{(\prime)} = A[1] \rtimes_{\alpha} H^{(\prime)} \subset \mathcal{G}$ given by
\begin{equation}
H \; = \; \text{Stab}_{\sigma}(1) \, , \qquad \lambda \; = \; \chi_1|_H \, , \qquad u \; = \; c_1|_H \, , \qquad p \; = \; s_1|_H \, ,
\end{equation}
and similarly for the $'$-ed variables. We then consider a fixed orbit of the product $G$-action $\sigma' \times \sigma$ on $[n'] \times [n]$ with fixed representative $(i_0,j_0) \in [n'] \times [n]$. As the $G$-action $\sigma$ on $[n]$ is transitive, we may without loss of generality assume that $j_0 =1$. Similarly, since the $G$-action $\sigma'$ on $[n']$ is transitive, we can fix $x \in G$ such that $x \triangleright 1 = i_0$\footnote{For fixed $i_0$, $x \in G$ is unique up to right multiplication by elements $h' \in \text{Stab}_{\sigma'}(1) \equiv H'$. Moreover, multiplying $x$ by elements $h \in \text{Stab}_{\sigma}(1) \equiv H$ from the left changes the representative $(i_0,1) \to (h \hspace{1pt}\triangleright \hspace{1pt} i_0,1)$ of the fixed $G$-orbit in $[n'] \times [n]$. The element $x \in G$ hence defines a double coset $[x] \in H \backslash G / H'$.}. Then, the stabiliser of the orbit representative $(i_0,1) \in [n'] \times [n]$ is given by
\begin{equation}
\begin{aligned}
\text{Stab}_{\sigma'\hspace{-0.5pt} \times \sigma}(i_0,1) \; &= \; \text{Stab}_{\sigma}(1) \, \cap \, \text{Stab}_{\sigma'}(i_0) \\
&= \; \text{Stab}_{\sigma}(1) \, \cap \, {}^x(\text{Stab}_{\sigma'}(1)) \; \equiv \; H \cap {}^{x\!}H' \; .
\end{aligned}
\end{equation}
Now let $\eta = (V,\varphi)$ be an intertwiner between $\rho$ and $\rho'$. Using the above, we can reduce the data associated to $\eta$ to the following:
\begin{itemize}
\item By defining $W:= V_{(i_0,1)}$, we obtain a finite-dimensional Hermitian space that vanishes unless $\chi'_{i_0} = \chi_1$. Since $\chi_1 \equiv \lambda$ and
\begin{equation}
\chi'_{i_0}(a) \; = \; \chi'_{x \hspace{1pt} \triangleright 1}(a) \; = \; \chi'_1(a^x) \; \equiv \; \lambda'(a^x) \; =: \; ({}^x\lambda')(a)
\end{equation}
for all $a \in A$, this means that $W = 0$ unless $\lambda = {}^x\lambda'$.
\item By defining for each $h \in H \cap {}^{x\!}H'$ the linear map
\begin{equation}
\psi(h) \; := \; \frac{c'_{i_0}(x,h^x)}{c'_{i_0}(h,x)} \, \cdot \, \varphi(h)_{(i_0,1)} \, : \;\; W \; \to \; W \; ,
\end{equation}
we obtain a projective representation $\psi$ of $H \cap {}^{x\!}H'$ on $W$ with projective 2-cocycle
\begin{equation}\label{eq-irr-intertwiner-2-cocycle}
\frac{{}^xu'}{u} \, \cdot \, \braket{\lambda,\gamma_x(\alpha)} \;\, \in \,\; Z^2\big(H \cap {}^{x\!}H',U(1)\big) \; ,
\end{equation}
where we defined the multiplicative factor
\begin{equation}
\gamma_x(\alpha)(h,k) \; := \; \frac{\alpha(h,x,k^x)}{\alpha(h,k,x) \cdot \alpha(x,h^x,k^x)} \; \in \; A \; . \label{eq-3d-intertwiner-2-cocycle}
\end{equation}
This representation then satisfies the conjugation rule
\begin{equation}  \label{eq-irr-intertwiner-conjugation-rule}
\psi(h)^{\dagger} \; = \; \Big(\frac{p}{{}^xp^{\hspace{0.5pt}\prime}}\Big)(h) \, \cdot \, \psi(h)^{-1} \; .
\end{equation}
\end{itemize}

Conversely, given $x \in G$ such that $\lambda = {}^x\lambda'$ together with a representation $\psi$ of $H \cap {}^{x\!}H'$ on a Hermitian space $W$ with projective 2-cocycle (\ref{eq-irr-intertwiner-2-cocycle}) and conjugation rule (\ref{eq-irr-intertwiner-conjugation-rule}), we obtain an intertwiner $\eta = (V,\varphi)$ between $\rho$ and $\rho'$ via induction: To this end, let $\lbrace r_1,...,r_n \rbrace$ and $\lbrace r'_1,...,r'_{n'} \rbrace$ be fixed sets of representatives of left $H$ and $H'$ cosets in $G$, i.e.
\begin{align}
G/H \; &= \; \lbrace r_1 H, ..., r_n H \rbrace \, , \\[2pt]
G/H' \; &= \; \lbrace r'_1 H, ..., r'_{n'} H \rbrace \, ,
\end{align}
such that $r_1 = r_1' = e$ and $r'_{i_0} = x$. As before, this allows us to define little group elements
\begin{align}
g_j \; &:= \; r_j^{-1} \cdot g \cdot r_{(g^{-1}) \hspace{0.7pt} \triangleright \hspace{0.7pt} j} \; \in \; H \\[2pt]
g'_i \; &:= \; (r'_i)^{-1} \cdot g \cdot r'_{(g^{-1}) \hspace{0.7pt} \triangleright \hspace{0.7pt} i} \; \in \; H'
\end{align}
for each $g \in G$ and all $i \in [n']$ and $j \in [n]$. We then define the double index set
\begin{equation}
I_x \; := \; \big\lbrace \hspace{1pt} (i,j) \in [n'] \!\times\! [n] \;\, \big| \;\, r_j^{-1} r_i' \, \in HxH' \hspace{1pt} \big\rbrace \; \subset \; [n'] \times [n]
\end{equation}
and fix for each $(i,j) \in I_x$ representatives $t_{ij} \in H$ and $t'_{ij} \in H'$ such that
\begin{equation}
r_j^{-1} r_i' \; = \; t_{ij} \cdot x \cdot (t_{ij}')^{-1}
\end{equation}
and $t_{i_0,1} = t'_{i_0,1} = 1$. Using this, we can construct for each $g \in G$ little group elements
\begin{equation}
\begin{aligned}
g_{ij} \; := \;\, &t^{-1}_{g \hspace{1pt} \triangleright \hspace{0.7pt} (ij)} \cdot g_{\hspace{1pt} g \hspace{1pt} \triangleright \hspace{0.7pt} j} \cdot t_{ij} \\[2pt] 
\equiv \,\; &{}^x\big[ \hspace{1pt} (t'_{g \hspace{1pt} \triangleright \hspace{0.7pt} (ij)})^{-1} \cdot g'_{\hspace{1pt} g \hspace{1pt} \triangleright \hspace{0.7pt} i} \cdot t'_{ij} \big] \;\, \in \; \, H \cap {}^{x\!}H' \; ,
\end{aligned}
\end{equation}
for all $(i,j) \in I_x$, which we can use to define the intertwiner $\eta = (V,\varphi)$ as follows:
\begin{itemize}
\item We define an $(n'\! \times \! n)$-matrix $V$ with Hermitian space entries 
\begin{equation}
V_{ij} \; := \; \begin{cases} W &\text{with} \, \braket{.,.}_{V_{ij}} \!:= \frac{p(t_{ij})}{p'(t'_{ij})} \hspace{-0.7pt}\cdot\hspace{-0.7pt} \braket{.,.}_W \; \text{if} \; (i,j) \in I_x\hspace{0.5pt}, \\
0 &\text{otherwise}. \end{cases} 
\end{equation}
\item For each $(i,j) \in I_x$ and $g \in G$, we construct a linear map $\varphi(g)_{ij}: V_{ij} \to V_{\hspace{1pt} g \hspace{1pt} \triangleright \hspace{0.7pt} (ij)}$ by
\begin{equation}\label{eq-induced-intertwiner}
\varphi(g)_{ij} \; := \; \frac{\nu_{ij}(u)(g)}{\nu'_{ij}(u')(g)} \, \cdot \, \Big\langle \lambda \hspace{1pt}, \, \frac{\mu_{ij}(\alpha)(g)}{{}^x[\hspace{0.5pt}\mu'_{ij}(\alpha)(g)\hspace{0.51pt}]} \, \cdot \, \omega_{x,ij}(\alpha)(g) \Big\rangle \, \cdot \, \psi(g_{ij}) \; ,
\end{equation}
where we defined the multiplicative phases
\begin{align}
\nu_{ij}(u)(g) \; &:= \; \frac{u\big( g_{ij}, t_{ij}^{-1}\big)}{u\big(t^{-1}_{g \hspace{1pt} \triangleright \hspace{0.7pt} (ij)}, \hspace{1pt} g_{\hspace{1pt} g \hspace{1pt} \triangleright \hspace{0.7pt} j}\big)} \; \in \; U(1) \; , \\[2pt]
\nu'_{ij}(u')(g) \; &:= \; \frac{u'\big( g_{ij}^{\hspace{1pt}x}, (t'_{ij})^{-1}\big)}{u'\big((t'_{g \hspace{1pt} \triangleright \hspace{0.7pt} (ij)})^{-1}, \hspace{1pt} g'_{\hspace{1pt} g \hspace{1pt} \triangleright \hspace{0.7pt} i}\big)} \; \in \; U(1) \; ,
\end{align}
as well as the multiplicative factors
\begin{align}
\mu_{ij}(\alpha)(g) \; &:= \; \frac{\alpha\!\left( t^{-1}_{g \hspace{1pt} \triangleright \hspace{0.7pt} (ij)}, r^{-1}_{\hspace{1pt} g \hspace{1pt} \triangleright \hspace{0.7pt} j}, g \right) \, \cdot \, \alpha\!\left( g_{ij}, t_{ij}^{-1}, r_j^{-1}\right)}{\alpha\!\left( t^{-1}_{g \hspace{1pt} \triangleright \hspace{0.7pt} (ij)}, g_{\hspace{1pt} g \hspace{1pt} \triangleright \hspace{0.7pt} j}, r_j^{-1} \right)} \; \in \; A \; , \\[5pt]
\mu'_{ij}(\alpha)(g) \; &:= \; \frac{\alpha\!\left( (t'_{g \hspace{1pt} \triangleright \hspace{0.7pt} (ij)})^{-1}, (r'_{\hspace{1pt} g \hspace{1pt} \triangleright \hspace{0.7pt} i})^{-1}, g \right) \, \cdot \, \alpha\!\left( g_{ij}^{\hspace{1pt} x}, (t'_{ij})^{-1}, (r'_i)^{-1}\right)}{\alpha\!\left( (t'_{g \hspace{1pt} \triangleright \hspace{0.7pt} (ij)})^{-1}, g'_{\hspace{1pt} g \hspace{1pt} \triangleright \hspace{0.7pt} i}, (r'_i)^{-1} \right)} \; \in \; A \; , \\[5pt]
\omega_{x,ij}(\alpha)(g) \; &:= \; \frac{\alpha\!\left( x, g_{ij}^{\hspace{1pt} x}, (r'_i \hspace{1pt} t'_{ij})^{-1} \right)}{\alpha\!\left( x, (r'_{\hspace{1pt} g \hspace{1pt} \triangleright \hspace{0.7pt} i} \hspace{1pt} t'_{g \hspace{1pt} \triangleright \hspace{0.7pt} (ij)})^{-1}, g \right) \, \cdot \, \alpha\!\left( g_{ij}, x, (r'_i \hspace{1pt} t'_{ij})^{-1} \right)} \; \in \; A \; .
\end{align}
The collection of linear maps (\ref{eq-induced-intertwiner}) then obeys
\begin{equation}
\varphi(g)_{h \, \triangleright \hspace{1pt} (ij)} \ \circ \, \varphi(h)_{ij} \,\; = \,\; \frac{c'_{gh \, \triangleright i}(g,h)}{c_{gh \, \triangleright j}(g,h)} \, \cdot \, \varphi(g \cdot h)_{ij} \; ,
\end{equation}
where $c \in C^2_{\sigma}(G,U(1)^n)$ and $c' \in C^2_{\sigma'}(G,U(1)^{n'})$ are as in (\ref{eq-induced-2-rep-2-cocycle}). Furthermore, it satisfies the conjugation rule
\begin{equation}
\varphi(g)_{ij}^{\dagger} \; = \; \frac{s_{g\, \triangleright j}(g)}{s'_{g\, \triangleright i}(g)} \, \cdot \, \varphi(g)_{ij}^{-1} \; ,
\end{equation}
where $s_j(g) \equiv p(g_j)$ and $s'_i(g) \equiv p'(g'_i)$.
\end{itemize}

To summarise, we can label the irreducible intertwiners between two irreducible unitary 2-representations $\rho=(H,\lambda,u,p)$ and $\rho'=(H',\lambda',u',p')$ by tuples consisting of the following pieces of data:
\begin{enumerate}
\item A representative $x \in G$ of a double coset $[x] \in H \backslash G / H'$ such that $\lambda = {}^x\lambda'$.
\item An irreducible representation $\psi$ of $H \cap {}^{x\!}H'$ with projective 2-cocycle
\begin{equation}
\frac{{}^xu'}{u} \, \cdot \, \braket{\lambda,\gamma_x(\alpha)} \;\, \in \,\; Z^2\big(H \cap {}^{x\!}H',U(1)\big)
\end{equation}
on an Hermitian space $W$ that satisfies the conjugation rule
\begin{equation}
\psi(h)^{\dagger} \; = \; \Big(\frac{p}{{}^xp^{\hspace{0.5pt}\prime}}\Big)(h) \, \cdot \, \psi(h)^{-1} \; .
\end{equation}
\end{enumerate}
We will write $\eta = (x,\psi)$ for an intertwiner specified by the above data in what follows. The identity morphism of an irreducible unitary 2-representation $\rho = (H,...)$ is the intertwiner $\text{id}_{\rho}: \rho \Rightarrow \rho$ with associated data $\text{id}_{\rho} = (e, \mathbf{1}_H)$, where $e \in G$ is the identity element and $\mathbf{1}_H$ is the trivial representation of the subgroup $H \subset G$. 

The duals and adjoints of an intertwiner $\eta = (x,\psi)$ between irreducible 2-representations $\rho = (H,\lambda,u,p)$ and $\rho' = (H',\lambda',u',p')$ can be described as follows:
\begin{itemize}
\item The \textit{1-dual} of $\eta$ is the intertwiner $\eta^{\vee_1}: (\rho')^{\vee_1} \Rightarrow \rho^{\vee_1}$ specified by $\eta^{\vee_1} = (x^{-1}, \psi^{\vee_1})$, where $\psi^{\vee_1}$ is the representation of $H' \cap H^x$ on $W$ defined by
\begin{equation}
\label{eq-irr-intertwiner-1-dual}
(\psi^{\vee_1})(k) \; = \; \frac{\psi({}^xk)}{\big\langle\lambda',\kappa_x(\alpha)(k)\big\rangle}
\end{equation}
with $\kappa_x(\alpha)(k) := \beta_{x^{-1},x}(\alpha)(k) \in A$ and $\beta(\alpha)$ as in (\ref{eq-composition-cochain}) below.
\item The \textit{2-dual} of $\eta$ is the intertwiner $\eta^{\vee_2}: \rho' \Rightarrow \rho$ specified by $\eta^{\vee_2} = (x^{-1}, \psi^{\vee_2})$, where $\psi^{\vee_2}$ is the representation of $H' \cap H^x$ on $W^{\vee}$ defined by
\begin{equation}
\label{eq-irr-intertwiner-2-dual}
(\psi^{\vee_2})(k) \; := \; \big\langle \lambda', \kappa_x(\alpha)(k) \big\rangle \, \cdot \, \big[\psi({}^xk)^{-1}\big]^{\vee} \; .
\end{equation}
Here, ${}^{\vee}$ denotes the transpose of linear maps between vector spaces. 
\item The \textit{adjoint} of $\eta$ is the intertwiner $\eta^{\dagger_1}: \rho' \Rightarrow \rho$ specified by $\eta^{\dagger_1} = (x^{-1},\psi^{\dagger_1})$, where $\psi^{\dagger_1}$ is the representation of $H' \cap H^x$ on $W^{\ast}$ defined by
\begin{equation}
\label{eq-irr-intertwiner-adjoint}
(\psi^{\dagger_1})(k) \; := \; \big\langle \lambda', \kappa_x(\alpha)(k) \big\rangle \, \cdot \, \psi({}^xk)^{\ast} \; .
\end{equation}
Here, ${}^{\ast}$ denotes the complex conjugate of linear maps between vector spaces.
\end{itemize}
In particular, if $\rho = \one$ is the trivial 2-representation and $\rho' = (H',1,u',p')$, then $\psi$ is given by a projective representation of $H'$ with 2-cocycle $u'$. In this case, $\psi^{\vee_2}$ and $\psi^{\dagger_1}$ correspond to the \textit{dual} and \textit{conjugate} representation of $\psi$, respectively.

\subsubsection{Composition} 

Given two intertwiners $\eta: \rho \Rightarrow \rho'$ and $\eta': \rho' \Rightarrow \rho''$ between three unitary 2-representations $\rho$, $\rho'$ and $\rho''$, we can compose them to obtain an intertwiner $\eta' \circ \eta: \rho \Rightarrow \rho''$. Concretely, if $\eta$ and $\eta'$ are specified by data $\eta = (V,\varphi)$ and $\eta'=(V',\varphi')$ as before, then their composition has associated data 
\begin{equation}\label{eq-intertwiner-composition}
(V',\varphi') \, \circ \, (V,\varphi) \; = \; \big( V' \boxtimes  V, \, \varphi' \boxtimes \varphi \big) \; ,
\end{equation}
where defined the matrix of Hermitian spaces and collection of linear maps
\begin{align}
(V' \boxtimes  V)_{ij} \;\, &= \;\, \bigoplus_{k\,=\,1}^{n'} \, V'_{ik} \hspace{1pt} \otimes \hspace{1pt} V_{kj} \; , \\[2pt]
(\varphi' \boxtimes  \varphi)(g)_{ij} \;\, &= \,\; \bigoplus_{k\,=\,1}^{n'} \, \varphi'(g)_{ik} \hspace{1pt} \otimes \hspace{1pt}  \varphi(g)_{kj} \; .
\end{align}

Now suppose that $\rho$, $\rho'$ and $\rho''$ are irreducible unitary 2-representations, so that we can label them by data $\rho = (H,\lambda,u,p)$ and similarly for $\rho'$ and $\rho''$. We furthermore assume that $\eta$ and $\eta'$ are irreducible intertwiners so that we can label them by data $\eta = (x,\psi)$ and $\eta' = (x',\psi')$ as before. Then, their composition is the (not necessarily irreducible) intertwiner that is labelled by the following data\footnote{In order to improve readability, we temporarily change the order in which we denote the composition of 1-morhisms, so that $(x,\psi) \circ (x',\psi')$ denotes the composition of $\eta: \rho \Rightarrow \rho'$ and $\eta': \rho' \Rightarrow \rho''$.}:
\begin{equation}
\label{eq-irr-intertwiner-composition}
\begin{gathered}
\quad (x,\psi) \, \circ \, (x',\psi') \;\; = \\[6pt]
\bigoplus_{[h] \, \in \, H^x \backslash \, H' / \, {}^{x'\hspace{-2pt}}H''} \left( \hspace{1pt} x \!\cdot\! h \!\cdot\! x', \; \text{Ind}_{H \, \cap \, {}^{x\!}H' \, \cap \, {}^{xhx'\hspace{-2pt}}H''}^{H \, \cap \, {}^{xhx'\hspace{-2pt}}H''}\!\left[ \, \dfrac{{}^{x}[\varepsilon_h(u')]}{\braket{\lambda, \beta_{x,h}(\alpha) \cdot \beta_{xh,x'}(\alpha)}} \cdot \big(\psi \otimes {}^{xh}\psi'\big) \, \right] \right) \; .
\end{gathered}
\end{equation}
Here, $\text{Ind}$ denotes the induction functor for (projective) representations of subgroups and we defined the 1-cochains 
\begin{align}
\varepsilon_h(u')(k) \,\; &:= \,\; \frac{u'(h,k^h)}{u'(k,h)} \,\; \in \,\; U(1) \; , \\[2pt]
\beta_{x,y}(\alpha)(k) \,\; &:= \,\; \frac{\alpha(k,x,y) \, \cdot \, \alpha(x,y,k^{xy})}{\alpha(x,k^x,y)} \,\; \in \; A \; . \label{eq-composition-cochain}
\end{align}
The above composition rule simplifies if we restrict attention to endomorphisms of an irreducible unitary 2-representation $\rho = (H,\lambda,u,p)$ with $H \subset G$ normal. In this case, irreducible endomorphisms $\eta = (x, \psi)$ and $\eta' =(x',\psi')$ are labelled by group elements $[x],[x'] \in G/H$ together with irreducible (projective) representations $\psi$ and $\psi'$ of $H$ and compose according to
\begin{equation}
(x,\psi) \, \circ \, (x',\psi') \; = \; \Bigg( \hspace{1pt} x \cdot x', \; \frac{\psi \otimes {}^x\psi'}{\big\langle \lambda, \beta_{x,x'}(\alpha)\big\rangle} \hspace{1pt} \Bigg) \; .
\end{equation}

\subsubsection{Equivalences}
\label{sssec-equivalences}

Having established the notion of intertwiners for unitary 2-representations, we can now discuss equivalences between them. Two unitary 2-representations $\rho = (n,\sigma,\chi,c,s)$ and $\rho' = (n',\sigma',\chi',c',s')$ are equivalent if there exist an invertible intertwiner $\eta: \rho \Rightarrow \rho'$ between them. Now let $\eta$ be specified by data $\eta = (V,\varphi)$ as before. Invertibility of $\eta$ can then be reduced to the following:
\begin{itemize}
\item As $V$ is an invertible $(n'\! \times \! n)$-matrix of Hermitian spaces, we must have that $n=n'$ with $V$ being of the form $V_{ij} = \delta_{i,\tau(j)} \cdot \mathbb{C}_{z_i}$ for some permutation $\tau \in S_n$ and some $z \in (\mathbb{Z}_2)^n$. Furthermore, since $V_{ij} = 0$ unless $\chi_i' = \chi_j$, we must have that $\chi' = {}^{\tau\!}\chi$, where $({}^{\tau\!}\chi)_i = \chi_{\tau^{-1}(i)}$.
\item As $\varphi$ provides isomorphisms $\varphi(g)_{ij}: V_{ij} \to V_{\sigma'_g(i),\sigma_g(j)}$ for each $g \in G$, we must have that $\sigma'_g = \tau \circ \sigma_g \circ \tau^{-1}$ for all $g \in G$. Furthermore, since the entries of $V$ are one-dimensional, the above linear maps need to be of the form 
\begin{equation}
\varphi(g)_{ij} \; = \; \delta_{i,\tau(j)} \cdot \vartheta_{g \hspace{1pt} \triangleright \hspace{0.7pt} i}(g)
\end{equation}
for some multiplicative phases $\vartheta_i(g) \in U(1)$. Plugging this into the composition rule (\ref{eq-intertwiner-composition-rule}) then yields
\begin{equation}
(d\vartheta)_i(g,h) \;\, \equiv \;\, \frac{\vartheta_{g^{-1} \hspace{1pt} \triangleright \hspace{0.7pt} i}(h) \cdot \vartheta_i(g)}{\vartheta_i(gh)} \;\, = \;\, \frac{c'_i(g,h)}{c_{\tau^{-1}(i)}(g,h)} \; ,
\end{equation}
which implies that $[c'/\,{}^{\tau\hspace{-1pt}}c] = 1 \in H^2_{\sigma'}(G,U(1)^n)$. In addition, the conjugation rule (\ref{eq-intertwiner-conjugation-rule}) yields
\begin{equation}
(dz)_i(g) \; \equiv \; \frac{z_{(g^{-1}) \hspace{0.7pt} \triangleright \hspace{0.7pt} i}}{z_i} \; = \; \frac{s'_i(g)}{s_{\tau^{-1}(i)}(g)} \; ,
\end{equation}
which implies that $[s'] = [{}^{\hspace{1pt}\tau\hspace{-1pt}}s] \in H^1_{\sigma'}(G,(\mathbb{Z}_2)^n)$.
\end{itemize}
Note that with $\eta=(V,\varphi)$ as above, the inverse of $\eta$ is given by its adjoint $\eta^{\dagger_1}=(V^{\dagger_1},\varphi^{\dagger_1})$. This shows that $\rho$ and $\rho'$ are in fact unitarily equivalent.

In summary, two unitary 2-representations $\rho = (n,\sigma,\chi,c,s)$ and $\rho' = (n',\sigma',\chi',c',s')$ are (unitarily) equivalent if they have the same dimension $n = n'$ and there exists a permutation $\tau \in S_n$ such that
\begin{equation}
\label{eq-intertwiner-equivalence}
\sigma' = {}^{\tau\!}\sigma \; , \qquad \chi' = {}^{\tau\!}\chi \; ,  \qquad [\hspace{0.5pt}c'/\,{}^{\tau\hspace{-1pt}}c\hspace{1pt}] = 1 \; , \qquad [s'] = [{}^{\hspace{1pt}\tau\hspace{-1pt}}s] \; .
\end{equation}
Now suppose that $\rho$ and $\rho'$ are irreducible unitary 2-representations, so that we can label them by data $\rho = (H,\lambda,u,p)$ and $\rho' = (H',\lambda',u',p')$ as before. By similar reasoning as above, $\rho$ and $\rho'$ are then (unitarily) equivalent if there exists an $x \in G$ such that
\begin{equation}
H' = {}^{x\!}H \; , \qquad \lambda' = {}^{x\!}\lambda \; , \qquad \left[ \frac{u'}{{}^{x}u} \cdot \braket{\lambda, \gamma_x(\alpha)} \right] = 1 \; , \qquad p' \, = \, {}^xp \; .
\end{equation}
Note that under the canonical 2-functor $\mathcal{F}: (H,\lambda,u,p) \mapsto (H,\lambda,u)$, this reproduces the known notion of equivalence of ordinary irreducible 2-representations. Two equivalent unitary 2-representations are hence equivalent as ordinary 2-representations as well. The converse, however, is not true, since equivalence as unitary 2-representations additionally requires the associated group homomorphisms $p'$ and ${}^xp$ to agree.

\subsection{Positivity}
\label{ssec-positivity}

Having classified unitary 2-representation of a finite 2-group $\mathcal{G}$, we now turn to the special case of \textit{positive} unitary 2-representations, which correspond to $\dagger$-2-functors
\begin{equation}
\rho: \; B\mathcal{G} \; \to \; \text{Mat}(\text{Hilb})
\end{equation} 
forming the 2-category $\text{2Rep}^{\dagger}_+(\mathcal{G})$. As most of the analysis is completely analogous, we will only highlight the main differences and discuss their consequences in what follows. 

To begin with, the classification of positive unitary 2-representations $\rho$ is identical to the one given in subsection \ref{ssec-classification}, so that any such $\rho$ can be labelled by quintuples $\rho = (n,\sigma,\chi,c,s)$ consisting of
\begin{enumerate}
\item a non-negative integer $n \in \mathbb{N}$, called the \textit{dimension} of the 2-representation,
\item a permutation action $\sigma: G \to S_n$ of $G$ on $[n] := \lbrace 1,...,n \rbrace$,
\item acollection of $n$ characters $\chi \in (A^{\vee})^n$ satisfying $\chi_{g \hspace{1pt} \triangleright \hspace{1pt} i}(a) = \chi_i(a^g)$,
\item a twisted 2-cochain $c \in C^2_{\sigma}(G,U(1)^n)$ satisfying $d_{\sigma}c = \braket{\chi,\alpha} $,
\item a twisted 1-cocycle $s \in Z^1_{\sigma}(G,(\mathbb{Z}_2)^n)$.
\end{enumerate}

Given two positive unitary 2-representations $\rho = (n,\sigma,\chi,c,s)$ and $\rho' = (n',\sigma',\chi',c',s')$, intertwiners between them can be classified in analogy to the analysis performed in subsection \ref{ssec-intertwiners}. Concretely, such intertwiners $\eta$ can be labelled by tuples $\eta = (V,\varphi)$ consisting of $(n'\times n)$-matrices $V$ and $\varphi$ of Hilbert spaces $V_{ij}$ and linear maps $\varphi(g)_{ij}: V_{ij} \to V_{g \, \triangleright (ij)}$ between them satisfying the composition rule (\ref{eq-intertwiner-composition-rule}). However, the conjugation rule (\ref{eq-intertwiner-conjugation-rule}) now implies that
\begin{equation}
\big\lVert \hspace{1pt} \varphi(g)_{ij} \hspace{-1pt} \cdot \hspace{-1pt} v \hspace{1pt} \big\rVert^2_{V_{g \, \triangleright (ij)}} \; = \;\; \frac{s_{g\, \triangleright j}(g)}{s'_{g\, \triangleright i}(g)} \, \cdot \, \lVert \hspace{0.5pt} v \hspace{0.5pt} \rVert^2_{V_{ij}} 
\end{equation}
for all $v \in V_{ij}$, which, in order to be compatible with the positive definiteness of the inner products on the Hilbert spaces $V_{ij}$, requires $s'_i(g) = s_j(g)$ for all $g \in G$ and $(i,j) \in [n'] \times [n]$ for which $V_{ij} \neq 0$. Note that, in this case, the invertible linear maps $\varphi(g)_{ij}$ are all unitary. 

To summarise, we can label intertwiners $\eta$ between two positive unitary 2-representations $\rho = (n,\sigma,\chi,c,s)$ and $\rho' = (n',\sigma',\chi',c',s')$ by tuples $\eta = (V,\varphi)$ consisting of
\begin{enumerate}
\item an $(n'\! \times \! n)$-matrix of Hilbert spaces $V_{ij}$ with $V_{ij} = 0$ unless $\chi_i' = \chi_j$ and $s'_i = s_j$,
\item for each $g \in G$ a collection of unitary linear maps $\varphi(g)_{ij}: V_{ij} \to V_{g \, \triangleright (ij)}$ satisfying
\begin{equation}
\varphi(g)_{h \, \triangleright \hspace{1pt} (ij)} \ \circ \, \varphi(h)_{ij} \,\; = \,\; \frac{c'_{gh \, \triangleright i}(g,h)}{c_{gh \, \triangleright j}(g,h)} \, \cdot \, \varphi(g \cdot h)_{ij} \; .
\end{equation}
\end{enumerate}
The duals and adjoints of such an intertwiner are as in (\ref{eq-intertwiner-1-dual}), (\ref{eq-intertwiner-2-dual}) and (\ref{eq-intertwiner-adjoint}), respectively. The composition of intertwiners is as in (\ref{eq-intertwiner-composition}). Moreover, asking $\eta$ to invertible reveals that two positive unitary 2-representations $\rho = (n,\sigma,\chi,c,s)$ and $\rho' = (n',\sigma',\chi',c',s')$ are equivalent if $n' = n$ and there exists a permutation $\tau \in S_n$ such that
\begin{equation}
\sigma' = {}^{\tau\!}\sigma \; , \qquad \chi' = {}^{\tau\!}\chi \; ,  \qquad [\hspace{0.5pt}c'/\,{}^{\tau\hspace{-1pt}}c\hspace{1pt}] = 1 \; , \qquad s' = {}^{\tau\hspace{-1.5pt}}s \; .
\end{equation}
Importantly, comparing with (\ref{eq-intertwiner-equivalence}), an equivalence between two positive unitary 2-repre-sentations $\rho$ and $\rho'$ requires $s'$ and ${}^{\tau\hspace{-1.5pt}}s$ to agree as 1-cocycles in $Z^1_{\sigma'}(G,(\mathbb{Z}_2)^n)$, and not just as cohomology classes in $H^1_{\sigma'}(G,(\mathbb{Z}_2)^n)$, as was the case for unitary (non-positive) 2-representations. In particular, this implies that when classifying irreducible positive unitary 2-representations by subgroups $H \subset G$ in analogy to subsection \ref{sssec-irreducible-2-reps}, we can only employ the isomorphism
\begin{equation}
\label{eq-twisted-1-cocycle-isomorphism}
Z^1\big(G,(\mathbb{Z}_2)^{G/H}\big) \; \cong \; C^1(G,\mathbb{Z}_2)^H \; ,
\end{equation}
where the group of (normalised) $H$-covariant 1-cochains on $G$ is given by 
\begin{equation}
C^1(G,\mathbb{Z}_2)^H \; := \; \big\lbrace \hspace{1pt} q: G \to \mathbb{Z}_2 \; \big| \; q(h\cdot g) = q(h) \cdot q(g) \;\; \forall \; h \in H, \; g\in G \big\rbrace \; .
\end{equation}
As a result, the irreducible positive unitary 2-representations of $\mathcal{G} = A[1] \rtimes_{\alpha} G$ can be labelled by quadruples $\rho = (H,\lambda,u,q)$ consisting of
\begin{enumerate}
\item a subgroup $H \subset G$,
\item a $H$-invariant character $\lambda \in A^{\vee}$,
\item a 2-cochain $u \in C^2(H,U(1))$ satisfying $du = \braket{\lambda,\alpha|_H}$,
\item a $H$-covariant 1-cochain $q \in C^1\hspace{-1pt}(G,\mathbb{Z}_2)^H$.
\end{enumerate}
Two such positive unitary 2-representations $\rho = (H,\lambda,u,q)$ and $\rho' = (H',\lambda',u',q')$ are considered equivalent if there exists an $x \in G$ such that
\begin{equation}
H' = {}^{x\!}H \; , \qquad \lambda' = {}^{x\!}\lambda \; , \qquad \left[ \frac{u'}{{}^{x\!}u} \cdot \braket{\lambda, \gamma_x(\alpha)} \right] = 1 \; , \qquad q' \, = \, {}^xq \; .
\end{equation}
The canonical 2-functor $\mathcal{E}: \text{2Rep}^{\dagger}_+(\mathcal{G}) \to \text{2Rep}^{\dagger}(\mathcal{G})$ sends
\begin{equation}
(H,\lambda,u,q) \; \mapsto \; (H,\lambda,u,q|_H)
\end{equation}
and is essentially surjective. Two equivalent positive unitary 2-representations are hence equivalent as unitary 2-representations as well. The converse, however, is not true, since equivalence as positive unitary 2-representations additionally requires the associated 1-cochains $q'$ and ${}^xq$ to agree.

The irreducible intertwiners between two given irreducible positive unitary 2-representa-tions $\rho=(H,\lambda,u,q)$ and $\rho'=(H',\lambda',u',q')$ can be classified in analogy to the analysis performed in subsection \ref{sssec-irreducible-intertwiners}. Concretely, irreducible intertwiners $\eta$ can be labelled by tuples $\eta = (x,\psi)$ consisting of
\begin{enumerate}
\item a representative $x \in G$ of a double coset $[x] \in H \backslash G / H'$ such that $\lambda = {}^x\lambda'$ and for all $g \in G$ it holds that
\begin{equation}
q(g) \; = \; \frac{q'(x^{-1}g)}{q'(x^{-1})} \; ,
\end{equation}
\item an irreducible unitary representation $\psi$ of $H \cap {}^{x\!}H'$ on a Hilbert space $W$ with projective 2-cocycle
\begin{equation}
\frac{{}^xu'}{u} \, \cdot \, \braket{\lambda,\gamma_x(\alpha)} \;\, \in \,\; Z^2\big(H \cap {}^{x\!}H',U(1)\big) \; .
\end{equation}
\end{enumerate}
The duals and adjoints of such an intertwiner are as in (\ref{eq-irr-intertwiner-1-dual}), (\ref{eq-irr-intertwiner-2-dual}) and (\ref{eq-irr-intertwiner-adjoint}), respectively. Their composition is as in (\ref{eq-irr-intertwiner-composition}).

\subsection{Example}
\label{ssec-example}

As an example, let us describe the 2-category of positive unitary 2-representations of the group $G = \mathbb{Z}_2$. Since $H^2(\mathbb{Z}_2,U(1))=1$, the simple objects of $\text{2Rep}^{\dagger}_+(\mathbb{Z}_2)$ are classified by subgroups $H \subset \mathbb{Z}_2$ together with $H$-covariant 1-cochains $q \in C^1(\mathbb{Z}_2,\mathbb{Z}_2)^H$. Using
\begin{equation}
C^1(\mathbb{Z}_2,\mathbb{Z}_2)^1 \; = \; C^1(\mathbb{Z}_2,\mathbb{Z}_2)^{\mathbb{Z}_2} \; = \; \mathbb{Z}_2 \; ,
\end{equation}
there are hence four simple objects up to unitary equivalence,
\begin{equation}
\vspace{-5pt}
\begin{gathered}
\includegraphics[height=2.15cm]{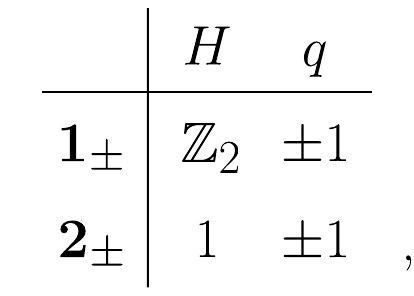}
\end{gathered}
\end{equation}
all of which are self-dual. Their morphism spaces can be determined to be
\begin{equation}
\begin{aligned}
\text{Hom}(\mathbf{1}_i,\mathbf{1}_j) \; &= \; \delta_{ij} \cdot \text{Rep}_+^{\dagger}(\mathbb{Z}_2) \; =: \; \braket{1,u_i} \, , \\
\text{Hom}(\mathbf{2}_i,\mathbf{2}_j) \; &= \; \delta_{ij} \cdot \text{Hilb}_{\hspace{1pt}\mathbb{Z}_2}\; =: \; \braket{1,v_i} \, , \\
\text{Hom}(\mathbf{1}_i,\mathbf{2}_j) \; &= \; \delta_{ij} \cdot\text{Hilb} \; =: \; \braket{x_i} \, , \\
\text{Hom}(\mathbf{2}_i,\mathbf{1}_j) \; &= \; \delta_{ij} \cdot\text{Hilb} \; =: \; \braket{y_i} \, ,
\end{aligned}
\end{equation}
which pictorially we illustrate as
\begin{equation}
\vspace{-5pt}
\begin{gathered}
\includegraphics[height=4cm]{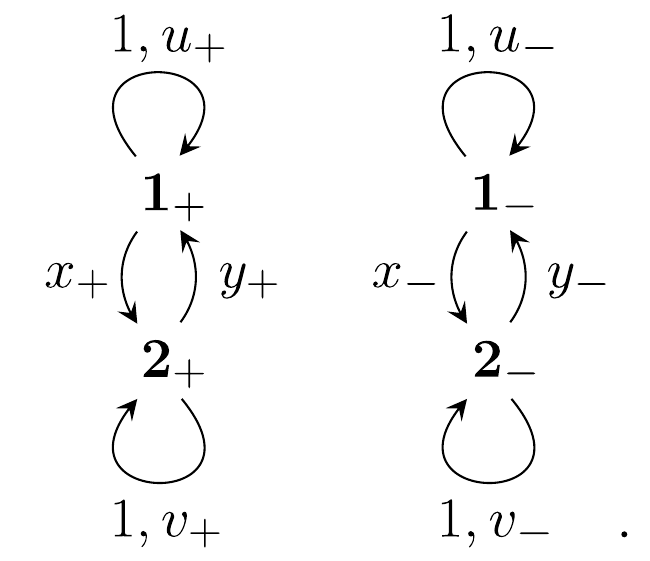}
\end{gathered}
\end{equation}
The composition rules of morphisms are given by
\begin{equation}
\begin{alignedat}{3}
(u_i)^2 \; &= \; 1 \, , &\qquad\quad x_i \, \circ \, u_i \; &= \; v_i \, \circ \, x_i \; = \; x_i  \, , &\qquad\quad x_i \, \circ \, y_i \; &= \; 1 \oplus v_i  \, ,  \\
(v_i)^2 \; &= \; 1  \, , &\qquad\quad y_i \, \circ \, v_i \; &= \; u_i \, \circ \, y_i \; = \; y_i  \, , &\qquad\quad y_i \, \circ \, x_i \; &= \; 1 \oplus u_i  \, . 
\end{alignedat}
\end{equation}
The duals and adjoints of morphisms are given by $(x_i)^{\vee} = (x_i)^{\dagger} = y_i$ with all other morphisms self-dual / self-adjoint. In particular, we see that, as a linear 2-category, $\text{2Rep}^{\dagger}_+(\mathbb{Z}_2)$ consists of two disjoint copies of $\text{2Rep}(\mathbb{Z}_2)$.

More generally, we can consider the image of $\text{2Rep}^{\dagger}_+(\mathbb{Z}_2)$ under the essentially surjective 2-functors $\mathcal{E}$ and $\mathcal{F}$, which act via
\begin{equation}
\vspace{-5pt}
\begin{gathered}
\includegraphics[height=3.1cm]{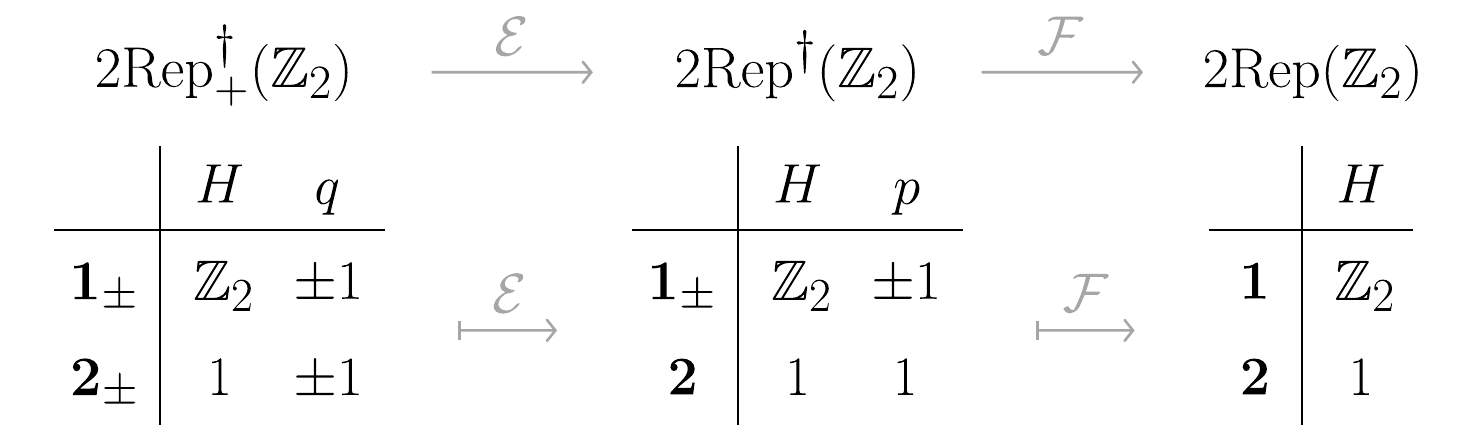}
\end{gathered}
\end{equation}
on simple objects and on morphisms via\footnote{For brevity, we only display morphisms in $\text{2Rep}^{\dagger}(\mathbb{Z}_2)$ up to isomorphism (not unitary isomorphism).}
\begin{equation}
\vspace{-5pt}
\begin{gathered}
\includegraphics[height=4.8cm]{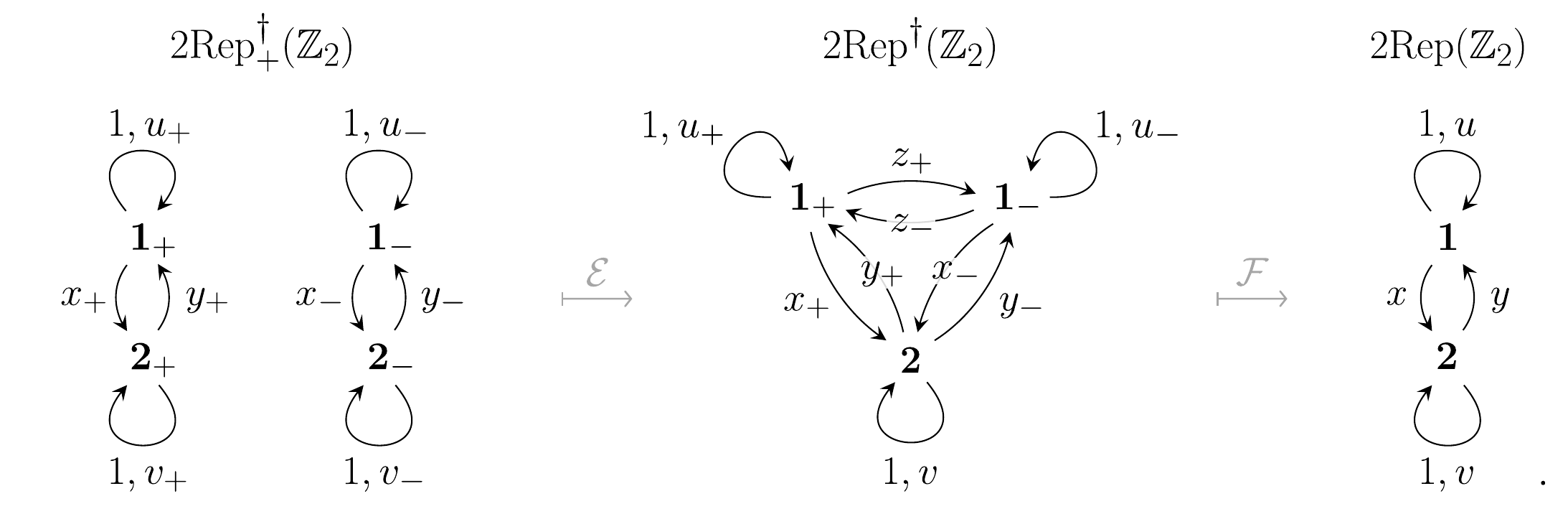}
\end{gathered}
\vspace{10pt}
\end{equation}
In particular, the 2-category $\text{2Rep}^{\dagger}(\mathbb{Z}_2)$ of unitary 2-representations of $\mathbb{Z}_2$ is connected due to the existence of additional morphisms $z_{\pm}$, whose composition rules up to isomorphism are given by 
\begin{equation}
\begin{alignedat}{3}
x_+ \,\circ\, z_- \; &= \; x_-^{\oplus 2} \, , &\qquad\qquad z_+ \, \circ \, z_- \; &= \; 1^{\oplus 2} \hspace{1pt}\oplus\hspace{1pt} u_-^{\oplus 2}  \, ,  \\
x_- \,\circ\, z_+ \; &= \; x_+^{\oplus 2} \, , &\qquad\qquad z_- \, \circ \, z_+ \; &= \; 1^{\oplus 2} \hspace{1pt}\oplus\hspace{1pt} u_+^{\oplus 2}  \, , \\[9pt]
z_+ \, \circ \, y_+ \; &= \; y_-^{\oplus 2} \, , &\qquad\qquad u_+ \, \circ \, z_- \; &= \; z_- \, \circ \, u_- \; = \; z_- \, , \\
z_- \, \circ \, y_- \; &= \; y_+^{\oplus 2} \, , &\qquad\qquad z_+ \, \circ \, u_+ \; &= \; u_- \, \circ \, z_+ \; = \; z_+ \, .
\end{alignedat}
\end{equation}
The duals and adjoints of $z_{\pm}$ are given by $(z_{\pm})^{\vee} = (z_{\pm})^{\dagger} = z_{\mp}$.

\vspace{8pt}
\textbf{Acknowledgements:} The author thanks Mathew Bullimore, Andrea Grigoletto and Tin Sulejmanpasic for helpful discussions. Thanks to Mathew Bullimore for useful comments on the draft.

\bibliographystyle{JHEP}
\bibliography{unitarity}

\end{document}